\documentclass[twocolumn, trackchanges]{aastex631}
\usepackage{amsmath}
\usepackage{graphicx}
\usepackage{epsf}
\usepackage{color}
\usepackage{enumitem}
\usepackage{csvsimple,booktabs}
\usepackage[flushleft]{threeparttable}
\usepackage{pifont}

\begin{document}

\title{\textbf{JADES: Detecting [OIII]$\lambda 4363$ Emitters and Testing Strong Line Calibrations in the High-\textit{z} Universe with Ultra-deep JWST/NIRSpec Spectroscopy up to $z \sim 9.5$}}

% Isaac L.
\author[0000-0003-4323-0597]{Isaac H. Laseter}
\affiliation{Department of Astronomy, University of Wisconsin -- Madison, Madison, WI 53706, USA}
% Michael M.
\author[0000-0003-0695-4414]{Michael V. Maseda}
\affiliation{Department of Astronomy, University of Wisconsin-Madison, Madison, WI 53706, USA}
% Mirko C.
\author[0000-0002-2678-2560]{Mirko Curti}
\affiliation{European Southern Observatory, Karl-Schwarzschild-Strasse 2, 85748 Garching, Germany}
\affiliation{Kavli Institute for Cosmology, University of Cambridge, Madingley Road, Cambridge, CB3 0HA, UK.}
\affiliation{Cavendish Laboratory - Astrophysics Group, University of Cambridge, 19 JJ Thomson Avenue, Cambridge, CB3 0HE, UK.}
% Roberto M.
\author[0000-0002-4985-3819]{Roberto Maiolino}
\affiliation{Kavli Institute for Cosmology, University of Cambridge, Madingley Road, Cambridge, CB3 0HA, UK.}
\affiliation{Cavendish Laboratory - Astrophysics Group, University of Cambridge, 19 JJ Thomson Avenue, Cambridge, CB3 0HE, UK.}
\affiliation{Department of Physics and Astronomy, University College London, Gower Street, London WC1E 6BT, UK.}
% Francesco D.
\author[0000-0003-2388-8172]{Francesco D'Eugenio}
\affiliation{Kavli Institute for Cosmology, University of Cambridge, Madingley Road, Cambridge, CB3 0HA, UK.}
\affiliation{Cavendish Laboratory - Astrophysics Group, University of Cambridge, 19 JJ Thomson Avenue, Cambridge, CB3 0HE, UK.}
% Alex C.
\author[0000-0002-0450-7306]{Alex J. Cameron}
\affiliation{Department of Physics, University of Oxford, Denys Wilkinson Building, Keble Road, Oxford OX1 3RH, UK}
% Tobias L.
\author[0000-0002-3642-2446]{Tobias J. Looser}
\affiliation{Kavli Institute for Cosmology, University of Cambridge, Madingley Road, Cambridge, CB3 0HA, UK.}
\affiliation{Cavendish Laboratory - Astrophysics Group, University of Cambridge, 19 JJ Thomson Avenue, Cambridge, CB3 0HE, UK.}
%%%%%%%%%%%%%%%%%%%
% Santiago A.
\author[0000-0001-7997-1640]{Santiago Arribas}
\affiliation{Centro de Astrobiolog\'ia (CAB), CSIC–INTA, Cra. de Ajalvir Km.~4, 28850- Torrej\'on de Ardoz, Madrid, Spain}
% William B.
\author[0000-0003-0215-1104]{William M. Baker}
\affiliation{Kavli Institute for Cosmology, University of Cambridge, Madingley Road, Cambridge, CB3 0HA, UK.}
\affiliation{Cavendish Laboratory - Astrophysics Group, University of Cambridge, 19 JJ Thomson Avenue, Cambridge, CB3 0HE, UK.}
% Rachana B.
\author[0000-0003-0883-2226]{Rachana Bhatawdekar}
\affiliation{European Space Agency (ESA), European Space Astronomy Centre (ESAC), Camino Bajo del Castillo s/n, 28692 Villanueva de la Cañada, Madrid, Spain; European Space Agency, ESA/ESTEC, Keplerlaan 1, 2201 AZ Noordwijk, NL}
% Kristan Boyett
\author[0000-0003-4109-304X]{Kristan Boyett}
\affiliation{School of Physics, University of Melbourne, Parkville 3010, VIC, Australia}
\affiliation{ARC Centre of Excellence for All Sky Astrophysics in 3 Dimensions (ASTRO 3D), Australia}
% Andrew B.
\author[0000-0002-8651-9879]{Andrew J.\ Bunker}
\affiliation{Department of Physics, University of Oxford, Denys Wilkinson Building, Keble Road, Oxford OX1 3RH, UK}
% Stefano C.
\author[0000-0002-6719-380X]{Stefano Carniani}
\affiliation{Scuola Normale Superiore, Piazza dei Cavalieri 7, I-56126 Pisa, Italy}
% Stephane C.
\author[0000-0003-3458-2275]{Stephane Charlot}
\affiliation{Sorbonne Universit\'e, CNRS, UMR 7095, Institut d'Astrophysique de Paris, 98 bis bd Arago, 75014 Paris, France}
% Jacopo C.
\author[0000-0002-7636-0534]{Jacopo Chevallard}
\affiliation{Department of Physics, University of Oxford, Denys Wilkinson Building, Keble Road, Oxford OX1 3RH, UK}
% Emma L.
\author[0000-0002-9551-0534]{Emma Curtis-lake}
\affiliation{Centre for Astrophysics Research, Department of Physics, Astronomy and Mathematics, University of Hertfordshire, Hatfield AL10 9AB, UK}
% Eiichi E.
\author[0000-0003-1344-9475]{Eiichi Egami}
\affiliation{Steward Observatory University of Arizona 933 N. Cherry Avenue Tucson AZ 85721, USA}
% Daniel E.
\author[0000-0002-2929-3121]{Daniel J.\ Eisenstein}
\affiliation{Center for Astrophysics $|$ Harvard \& Smithsonian, 60 Garden St., Cambridge MA 02138 USA}
% Kevin H.
\author[0000-0003-4565-8239]{Kevin Hainline}
\affiliation{Steward Observatory University of Arizona 933 N. Cherry Avenue Tucson AZ 85721 USA}
% Ryan H.
\author[0000-0002-8543-761X]{Ryan Hausen}
\affiliation{Department of Physics and Astronomy, The Johns Hopkins University, 3400 N. Charles St., Baltimore, MD 21218}
% Zhiyuan J.
\author[0000-0001-7673-2257]{Zhiyuan Ji}
\affiliation{Steward Observatory University of Arizona 933 N. Cherry Avenue Tucson AZ 85721, USA}
% Nimisha Kumari
\author[0000-0002-5320-2568]{Nimisha Kumari}
\affiliation{AURA for European Space Agency, Space Telescope Science Institute, 3700 San Martin Drive. Baltimore, MD, 21210}
% Michele P.
\author[0000-0002-0362-5941]{Michele Perna}
\affiliation{Centro de Astrobiolog\'ia (CAB), CSIC–INTA, Cra. de Ajalvir Km.~4, 28850- Torrej\'on de Ardoz, Madrid, Spain}
%Tim R.
\author[0000-0002-7028-5588]{Tim Rawle}
\affiliation{European Space Agency (ESA), European Space Astronomy Centre (ESAC), Camino Bajo del Castillo s/n, 28692 Villafranca del Castillo, Madrid, Spain}
%Hans-Walter R.
\author[0000-0003-4996-9069]{Hans-Walter Rix}
\affiliation{Max-Planck-Institut f\"ur Astronomie, 
K\"onigstuhl 17, D-69117, Heidelberg, Germany}
% Brant R.
\author[0000-0002-4271-0364]{Brant Robertson}
\affiliation{Department of Astronomy and Astrophysics University of California, Santa Cruz, 1156 High Street, Santa Cruz CA 96054, USA \affiliation{Department of Astronomy and Astrophysics, University of California, Santa Cruz, 1156 High Street, Santa Cruz, CA 95064, USA}}
% Bruno R.
\author[0000-0001-5171-3930]{Bruno Rodríguez Del Pino}
\affiliation{Centro de Astrobiolog\'ia (CAB), CSIC–INTA, Cra. de Ajalvir Km.~4, 28850- Torrej\'on de Ardoz, Madrid, Spain}
% Lester S.
\author[0000-0001-9276-7062]{Lester Sandles}
\affiliation{Kavli Institute for Cosmology, University of Cambridge, Madingley Road, Cambridge, CB3 0HA, UK.}
\affiliation{Cavendish Laboratory - Astrophysics Group, University of Cambridge, 19 JJ Thomson Avenue, Cambridge, CB3 0HE, UK.}
% Jan Sc.
\author{Jan Scholtz}
\affiliation{Kavli Institute for Cosmology, University of Cambridge, Madingley Road, Cambridge, CB3 OHA, UK.}
\affiliation{Cavendish Laboratory - Astrophysics Group, University of Cambridge, 19 JJ Thomson Avenue, Cambridge, CB3 OHE, UK.}

% Renske S.
\author[0000-0001-8034-7802]{Renske Smit}
\affiliation{Astrophysics Research Institute, Liverpool John Moores University, 146 Brownlow Hill, Liverpool L3 5RF, UK}
% Sandro Tacchella
\author[0000-0002-8224-4505]{Sandro Tacchella}
\affiliation{Kavli Institute for Cosmology, University of Cambridge, Madingley Road, Cambridge, CB3 0HA, UK.}
\affiliation{Cavendish Laboratory - Astrophysics Group, University of Cambridge, 19 JJ Thomson Avenue, Cambridge, CB3 0HE, UK.}
% Hannah U.
\author[0000-0003-4891-0794]{Hannah \"Ubler}
\affiliation{Kavli Institute for Cosmology, University of Cambridge, Madingley Road, Cambridge, CB3 0HA, UK.}
\affiliation{Cavendish Laboratory - Astrophysics Group, University of Cambridge, 19 JJ Thomson Avenue, Cambridge, CB3 0HE, UK.}
% Christina W.
\author[0000-0003-2919-7495]{Christina C. Williams}
\affiliation{NSF’s National Optical-Infrared Astronomy Research Laboratory, 950 North Cherry Avenue, Tucson, AZ 85719, USA}
\affiliation{Steward Observatory University of Arizona 933 N. Cherry Avenue Tucson AZ 85721, USA}
% Chris Willott
\author[0000-0002-4201-7367]{Chris Willott}
\affiliation{NRC Herzberg, 5071 West Saanich Rd, Victoria, BC V9E 2E7, Canada}
% Joris W
\author[0000-0002-7595-121X]{Joris Witstok}
\affiliation{Kavli Institute for Cosmology, University of Cambridge, Madingley Road, Cambridge, CB3 0HA, UK.}
\affiliation{Cavendish Laboratory - Astrophysics Group, University of Cambridge, 19 JJ Thomson Avenue, Cambridge, CB3 0HE, UK.}

%%%%%%%%%%%%%%%%%% Abstract %%%%%%%%%%%%%%%%%%

\begin{abstract}   
We present 10 novel [OIII]$\lambda 4363$ auroral line detections up to $z\sim 9.5$ measured from ultra-deep JWST/NIRSpec MSA spectroscopy from the JWST Advanced Deep Extragalactic Survey (JADES). We leverage the deepest spectroscopic observations yet taken with NIRSpec to determine electron temperatures and oxygen abundances using the direct T$_e$ method. We directly compare against a suite of locally calibrated strong-line diagnostics and recent high-\textit{z} calibrations. We find the calibrations fail to simultaneously match our JADES sample, thus warranting a \textit{self-consistent} revision of these calibrations for the high-\textit{z} Universe. We find weak dependence between R2 and O3O2 with metallicity, thus suggesting these line-ratios are ineffective in the high-\textit{z} Universe as metallicity diagnostics and degeneracy breakers. We find R3 and R23 still correlate with metallicity, but we find tentative flattening of these diagnostics, thus suggesting future difficulties when applying these strong-line ratios as metallicity indicators in the high-\textit{z} Universe. We also propose and test an alternative diagnostic based on a different combination of R3 and R2 with a higher dynamic range. We find a reasonably good agreement (median offset of 0.002 dex, median absolute offset of 0.13 dex) with the JWST sample at low metallicity, but future investigation is required on larger samples to probe past the turnover point. At a given metallicity, our sample demonstrates higher ionization/excitation ratios than local galaxies with rest-frame EWs(H$\beta$) $\approx 200 -300$~\AA. However, we find the median rest-frame EWs(H$\beta$) of our sample to be $\sim 2\text{x}$ less than the galaxies used for the local calibrations. This EW discrepancy combined with the high ionization of our galaxies does not present a clear description of [OIII]$\lambda 4363$ production in the high-\textit{z} Universe, thus warranting a much deeper examination into the factors affecting production.

\end{abstract}

%%%%%%%%%%%%%%%%%% Introduction %%%%%%%%%%%%%%%%%%

\section{\textbf{Introduction}} \label{Introduction}

Before the era of the James Webb Space Telescope (JWST), our understanding of the interstellar medium (ISM) of high redshift ($z \gtrsim 3$) galaxies was limited to identifying potential local analogs, such as extremely metal-poor galaxies (XMPGs) \citep{Izotov_2006, Izotov_2021, Laseter_2022, Thuan_2022},  extreme star-forming galaxies (e.g., blueberries \citep{Yang_2017a}, blue compact dwarf galaxies \citep{Sargent_1970, Cairos_2010C}, and green peas \citep{Cardamone_2009, Jaskot_2013, Henry_2015, Yang_2017b}), and damped Lyman-$\alpha$ systems \citep{Wolfe_2005}. Several ISM properties such as chemical abundances, ionization states, temperatures, and densities, which can reveal the sources powering the ionization and key evolutionary processes, can be probed by studying the ratio between different rest-frame optical emission lines such as [OII]$\lambda\lambda 3727, 3729$, [OIII]$\lambda\lambda 4959, 5007$ and the Hydrogen Balmer series. However, by $z \sim 3$, H$\alpha$ is unobservable from ground-based telescopes, and weaker lines are impractical to observe. Insights from rest-frame optical emission lines primarily originated from photometric techniques \citep[e.g.,][]{Shim_2011, Gonzalez_2012, Labbe_2013, Smit_2014, Rasappu_2016, Roberts-Borsani_2016}, but there were difficulties targeting faint sources (e.g., $\rm M_{UV} \approx -17$) that are known to exist at these redshifts from Lyman-$\alpha$ surveys \citep[e.g.,][]{Cowie_1998, Finkelstein_2007, Cowie_2011, Matthee_2015, Finkelstein_2016, Bacon_2017, Maseda_2018, Maseda2020, Taylor_2020, Taylor_2021, Reddy_2022, Wold_2022}. However, these limitations were alleviated when the JWST Early Release Observations (ERO) of SMACS $\text{J}0723.3-7327$ demonstrated clear observations of rest-frame optical emission lines \citep[e.g.,][]{Carnall_2023}, thus ushering in a new era of high-\textit{z} spectroscopic studies. 
 
One of these JWST observed rest-frame optical emission lines was [OIII]$\lambda4363$. [OIII]$\lambda4363$ is a so-called \textit{auroral line}, which are collisionally excited emission lines originating from higher energy levels compared to the typical nebular lines observed in galaxy spectra. Auroral lines are emitted by different ionic species and at different wavelengths (e.g., OIII]$\lambda\lambda 1661,1666$, [OIII]$\lambda 4363$, [OII]$\lambda\lambda 7320,7330$, [SII]$\lambda 4069$, [NII]$\lambda 5755$, and [SIII]$\lambda 6312$) \citep{Castellanos_2002, Maiolino_2019}. However, [OIII]$\lambda4363$ has become the most sought-after due to its strength compared to other auroral lines and its proximity to rest-frame optical emission lines. If observed, the ratio of [OIII]$\lambda4363$ to the stronger, lower energy level lines of [OIII]$\lambda\lambda 4959, 5007$ acts as an exceptional electron temperature diagnostic. If the electron temperature can be determined then gas-phase ionic abundances, i.e., \textit{metallicities}, can be derived directly from the strengths of common emission lines. This method of determining electron temperatures/metallicities is known as the ``direct $T_e$ method” ($T_e$) due to the direct comparison of energy levels of a single species. The main disadvantage of employing $T_e$ is the intrinsic faintness of [OIII]$\lambda4363$, which can be 10-100 times fainter than the neighboring oxygen and Balmer lines \citep{Maiolino_2019}. As such, observations of [OIII]$\lambda4363$ have been restricted predominately to low-z, low metallicity individual galaxies or to stacked spectra of several hundreds of galaxies \citep[e.g.,][]{Izotov_2006, Hirschauer_2016, Curti_2017, Hsyu_2017, Izotov_2021, Aver_2022, Laseter_2022}, with sparse detections at $z \geq 1$ \citep[e.g.,][]{Christensen_2012, Maseda_2014, Patricio_2018}, thus limiting our measurements of galaxy metallicities in the high-\textit{z} Universe. 

Measuring gas-phase metallicities is vital: Metallicity is sensitive to many physical processes driving the baryon cycle in galaxies as it is the result of the complex interplay between gas flows, star formation, and ISM enrichment \citep{Matteucci_2012, Maiolino_2019}. Massive effort has been committed to modeling the chemical evolution of galaxies and their surroundings to provide information into the relative importance of such processes. However, such models require tight observational constraints, which can be established by investigating the metallicity over cosmic time. At $z = 0$, there is a well constrained relationship between stellar masses and metallicity known as the mass-metallicity relation (MZR) \citep{Tremonti_2004, Kewley_2008, Mannucci_2010}. Evolution in the MZR has been shown to exist up to  $z \sim 3$ in the sense that galaxies at higher \textit{z} have lower metallicity at a given stellar mass. However, statistical studies of the MZR based on large samples of galaxies do not typically determine metallicities by the direct $T_e$ method due to the difficulties in detecting [OIII]$\lambda 4363$, especially at higher \textit{z} and in higher metallicity galaxies. Most studies derive metallicities through \textit{strong-line diagnostics}. 

Strong-line calibrations typically exploit optical nebular lines (e.g., [OIII]$\lambda 5007$, [NII]$\lambda 6584$, [SII]$\lambda 6717$, H$\beta$, etc.) that are calibrated against metallicities derived through the direct $T_e$ method \citep[e.g.,][]{Curti_2017, Curti_2020, Bian_2018, Sanders_2021, Nakajima_2022}, with photoionization models \citep[e.g.,][]{Perez_2014, Dopita_2016}, or a hybrid combination of the two \citep[e.g.,][]{Pettini_2004, Tremonti_2004, Maiolino_2008}. However, it has been shown that even for the same galaxy population different calibrations can disagree by up to 0.6 dex \citep{Kewley_2008}. \cite{Curti_2017, Curti_2020} improved calibrations by stacking Sloan Digital Sky Survey (SDSS) galaxies to provide a full empirical calibration for a suite of optical nebular emission lines. However, the properties of the high \textit{z} universe differ from the local universe, so it is highly uncertain whether locally calibrated strong line diagnostics are appropriate to use in the early Universe.

The pivotal change in this predicament is the observational ability of JWST combined with the near-infrared spectrograph NIRSpec \citep{Boker_2022,Jakobsen_2022, Ferruit_2022, Boker_2023}. NIRSpec has opened the capability of obtaining multi-object spectroscopy in the near-IR from space with unmatched sensitivity compared to any current or past facility. JWST/NIRSpec has already observed a number of [OIII]$\lambda 4363$ emitters \citep[e.g.,][]{Schaerer_2022, Taylor_2022, Curti_2023, Trump_2023, Rhoads_2023}, though all these previous works were based on observations from Early Release Observations (ERO) data obtained by targeting galaxies lensed by the cluster SMACS J0723.3-7327 \citep{Repp_2018} and a number of extraction and metallicity prescriptions were employed. Recently, \cite{Nakajima_2023} reanalyzed 4 sources from ERO and 4 sources from GLASS, along with identifying a new [OIII]$\lambda 4363$ source from CEERS in the EGS. \cite{Sanders_2023} also identified 16 galaxies with [OIII]$\lambda 4363$ detections from CEERS. In addition, \cite{Ubler_2023} identified [OIII]$\lambda 4363$ in a low metallicity AGN at $z\sim 5.55$ with the JWST/NIRSpec Integral Field Spectrograph.

However, all of these observations were obtained with relatively shallow spectroscopy. For example, the CEERS observations across 6 pointings totaled $\sim 5$ hours of integration \citep{Finkelstein_2022} and the ERO observations across 2 pointings totaled $\sim 5$ hours of integration \citep{Carnall_2023}. Here we utilize deep spectroscopic data taken from the JWST Advanced Deep Extragalactic Survey (JADES), the deepest spectroscopic observations yet taken with NIRSpec, to provide a more detailed look at [OIII]$\lambda 4363$ detections and assess locally derived strong line calibrations up to $z \sim 9.5$. These NIRSpec/JADES observations obtained exposure times of up to 28 hours in the PRISM/CLEAR (R$\sim 100$) and up to 7 hours in each of the 3 medium resolution gratings (R$\sim1000$) and the G395H/F290L high resolution grating (R$\sim2700$), providing unprecedented new insights into chemical evolution and ISM properties of galaxies within the first Gyr of the Universe’s history.

The structure of this paper is as follows: In Section \ref{NIRSpec} we describe the JADES observations, data reduction and emission line flux measurements; in Section \ref{4363 Detections} we present our [OIII]$\lambda 4363$ detections; in Section \ref{Strong Line Calibrations} we compare our direct metallicity measurements to strong line calibrations calibrations; in Section \ref{Discussion} we discuss our findings; and finally in Section \ref{Conclusions} we present our conclusions. For this work we adopt the \cite{Planck_2020} cosmology: H$_0$ = 67.36 km/s/Mpc, $\rm \Omega_{m} = 0.3153$, $\rm \Omega_{\lambda}$ $= 0.6847$. 

%%%%%%%%%%%%%%%%%%%%%%%%%%%%%%%%%%%%%%%%%%%%%%%%%%%%%%%%%

\section{\textbf{Observations, Data Processing, and Data Analysis}} \label{NIRSpec}

\subsection{\textbf{Observations}} \label{Observations}

The data presented in this paper were obtained via multi-object spectroscopic observations from \emph{JWST}/NIRSpec using the micro-shutter assembly (MSA). Observations were carried out in three visits between Oct 21-25, 2022 (Program ID: 1210; PI: N. Luetzgendorf) in the Great Observatories Origins Deep Survey South (GOODS-S) legacy field as part of JADES. Each visit consisted of 33,613 s integration in the PRISM/CLEAR low-resolution setting and 8,403 s integration in each of G140M/F070LP, G235M/F170LP, G395M/F290LP, and G395H/F290LP filter/grating settings. Across three visits, this totals 28 hours of integration in the PRISM, which provides continuous spectral coverage from 0.6 - 5.3 \micron\, at $R\sim30-300$, and $\sim$ 7 hours in each of the medium resolution gratings, which combine to provide $R\sim700-1300$ across the full spectral range of NIRSpec, plus 7 hours in the high-resolution grating which provides $R\sim2700$ from $\sim$2.8 - 5.1 \micron~, though the exact wavelength coverage depends on the target location in the MSA.

Observations within each visit were performed as a 3-shutter nod. The central pointing of each visit was dithered (by $<1$ arcsec) such that common targets were observed in different shutters and different detector real-estate. Thus, each visit had a unique MSA configuration, although target allocation (performed with the eMPT \footnote{\url{https://github.com/esdc-esac-esa-int/eMPT_v1}}; \cite{Bonaventura_2023}) was optimised for maximising target commonality between all three dither positions.

A total of 253 unique targets were observed in the PRISM configuration with the three dithers featuring 145, 155, and 149 targets respectively. All targets are observed with non-overlapping spectra in the PRISM mode. However, in the medium and high resolution gratings, individual spectra are dispersed over a larger number of detector pixels, and thus there is a possibility of spectral overlap. To minimize contamination overlap, we isolate our highest priority targets by closing the shutters of low-priority targets on the same row (i.e. targets that would cause overlapping spectra) during observations. Thus, for our grating spectra we observe 198 unique targets (119, 121, and 111 in each dither).

\begin{figure}
    \centering
    \includegraphics[width = \columnwidth]{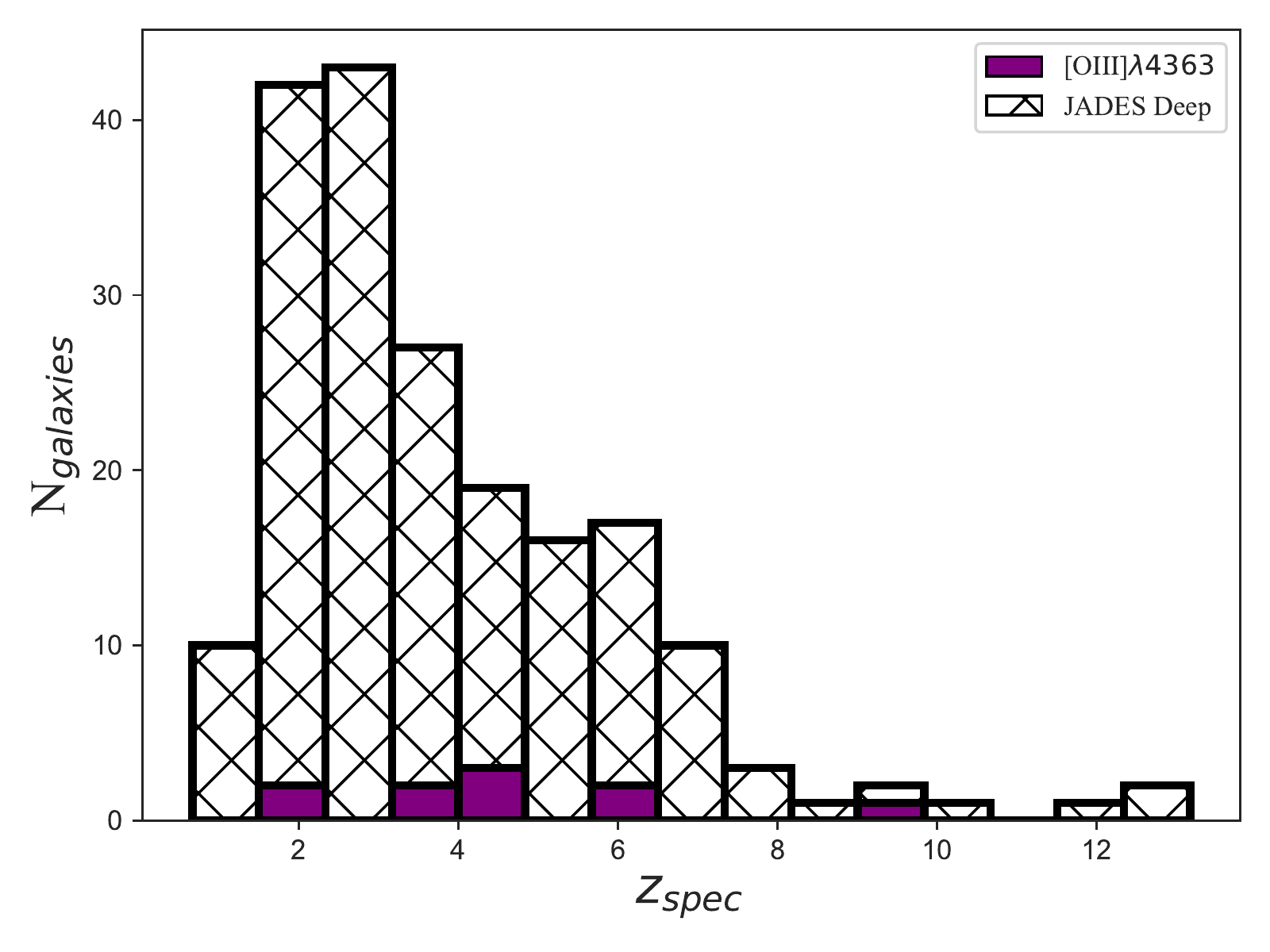}
    \caption{Redshift distribution of our parent JADES sample of 198 objects with both PRISM and grating spectra and our 10 novel [OIII]$\lambda 4363$ emitters. No pre-selection was performed on the parent JADES sample as we visually inspected all objects.}
    \label{fig: Histogram}
\end{figure}

\subsection{\textbf{Data Processing}} \label{Data Processing}

The JWST/NIRSpec observations have been processed by adopting algorithms developed by the  ESA NIRSpec Science Operations Team (SOT) and the NIRSpec GTO Team, and the details of the data-processing workflow will be presented a the forthcoming NIRSpec GTO collaboration paper. Once we retrieved the level-1a data from the MAST archive, we estimated the count rate per pixel by using the unsaturated groups in the ramp and removing jumps due to cosmic rays identified  by estimating the slope of the individual ramps. During this first stage, we also performed the master bias and dark subtraction, corrected snowball artifacts, and flagged saturated pixels.

We then performed the pixel-by-pixel background subtraction by combining the three nod exposures of each pointing. We note that for some targets we excluded one of the 3-shutter nods in the background subtraction stage as a  serendipitous source contaminated the open shutters.  We then created 2D dimensional (2D) cutouts of each 3-shutter slit and performed the flat-field, spectrograph optics, and dispersers corrections. Then we run the absolute calibration stage and corrected the 2D spectra for the path-losses depending on the relative position of the source within its shutter. We computed and applied the path-losses correction for a point-like source as the size of our targets are smaller or comparable to the spatial angular resolution of the telescope at the redshifted wavelength of the optical nebular lines at $z>7$. 

We rectified and interpolated the 2D continuum map onto a regular grid for all medium/high-resolution gratings and an irregular grid for the PRISM/CLEAR to avoid an oversampling of the line spread function at short wavelengths. Finally, the 1D spectra were extracted from the 2D map adopting a box-car aperture as large as the shutter size and centered on the relative position of the target in the shutter. For each target, we combined all 1D spectra and removed bad pixels by adopting a sigma-clipping approach. 

\subsection{\textbf{PPXF}} \label{PPXF}

Emission-line measurements and continuum modelling are made simultaneously using the penalised pixel fitting algorithm, \texttt{ppxf} \citep{Cappellari_2017, Cappellari_2022}.
\texttt{ppxf} models the continuum as a linear superposition of simple stellar-population (SSP) spectra, using
non-negative weights and matching the spectral resolution of the observed spectrum. As input, we used the high-resolution (R=10,000) SSP library combining MIST isochrones \citep{Choi_2016} and the C3K theoretical atmospheres \citep{Conroy_2018}. The flux blue-ward of the Lyman break was manually set to 0. These templates are complemented by a 5\textsuperscript{th}-degree multiplicative Legendre polynomial, to take into account systematic differences between the SSPs and the data (e.g., dust, mismatch between the SSP models and high-redshift stellar populations, and residual flux calibration problems). The emission lines are modelled as pixel-integrated Gaussians, again matching the observed spectral resolution. To reduce the number of degrees of freedom, we divide all emission lines in four kinematic groups, constrained to have the same redshift and \emph{intrinsic} broadening. These are UV lines (blueward of 3000~\AA), the Balmer series of Hydrogen, non-Hydrogen optical lines (blueward of 9000~\AA), and NIR lines. The stellar component has the same kinematics as the Balmer lines. Furthermore, we tie together doublets that have fixed ratios, and constrain variable-ratio doublets to their physical ranges. In particular, we fit for the following lines of interest: [OII]$\lambda\lambda 3726, 3729$ , [Ne III]$\lambda\lambda 3869, 3967$, H$\delta$, H$\gamma$, [OIII]$\lambda 4363$, H$\beta$, [OIII]$\lambda\lambda 4959, 5007$, H$\alpha$, [NII]$\lambda 6583$, [SII]$\lambda\lambda 6716, 6731$.

\section{\textbf{[OIII]$\lambda$4363 detections and the $T_e$ method}} \label{4363 Detections}

\subsection{\textbf{JADES}}

\begin{figure*}[hbt!]
    \centering
    \includegraphics[width = \columnwidth]{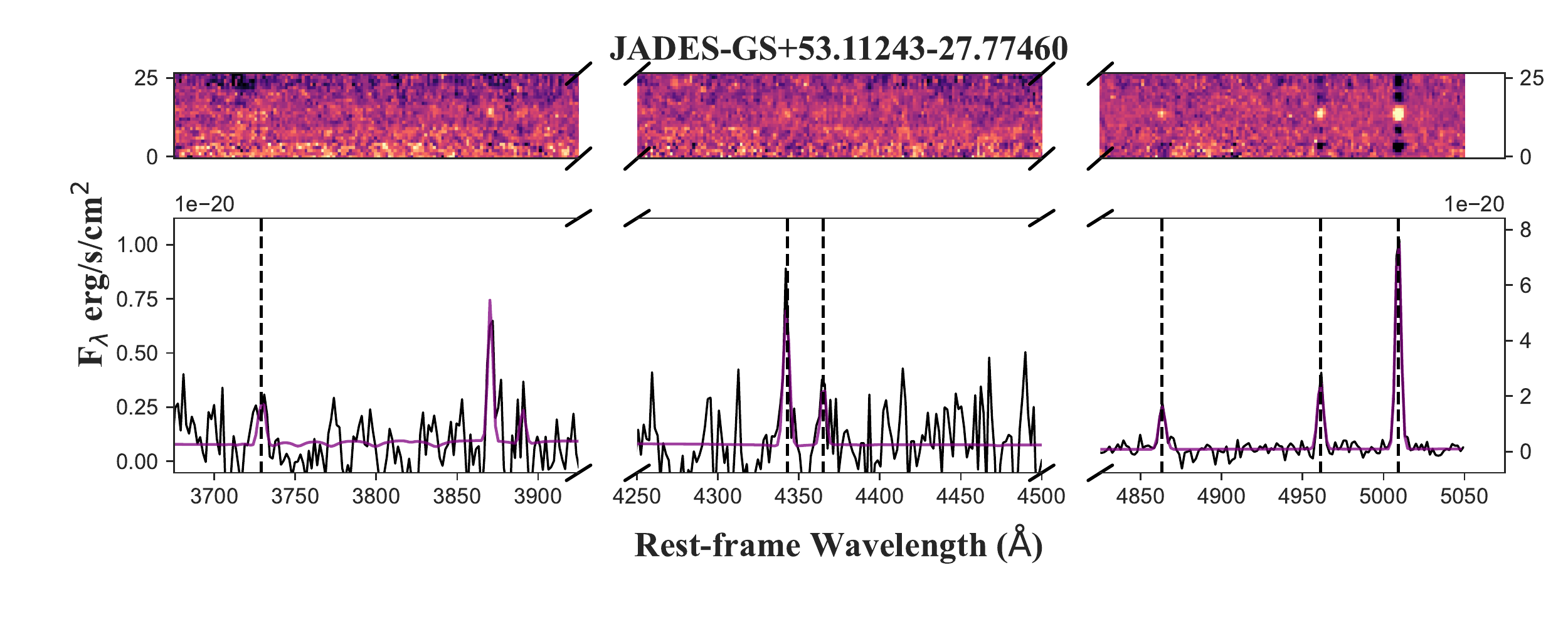}
    \includegraphics[width = \columnwidth]{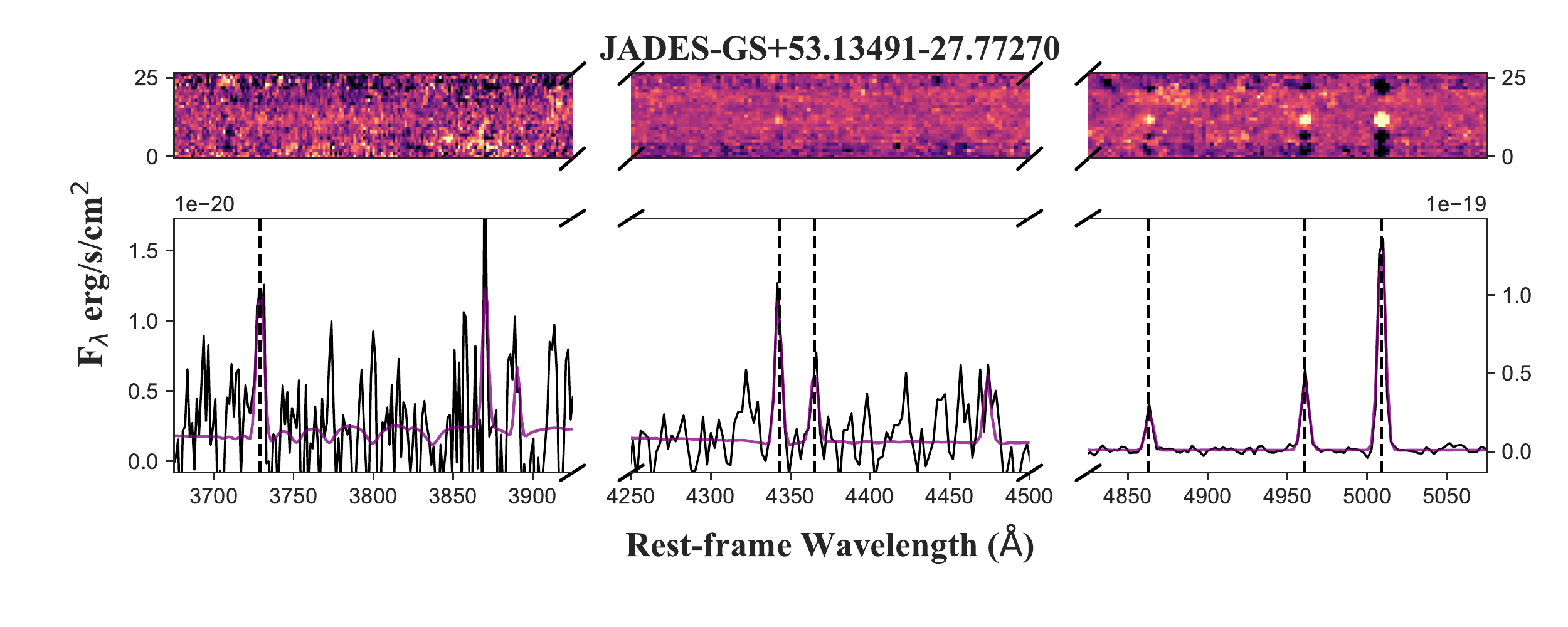}
    \includegraphics[width = \columnwidth]{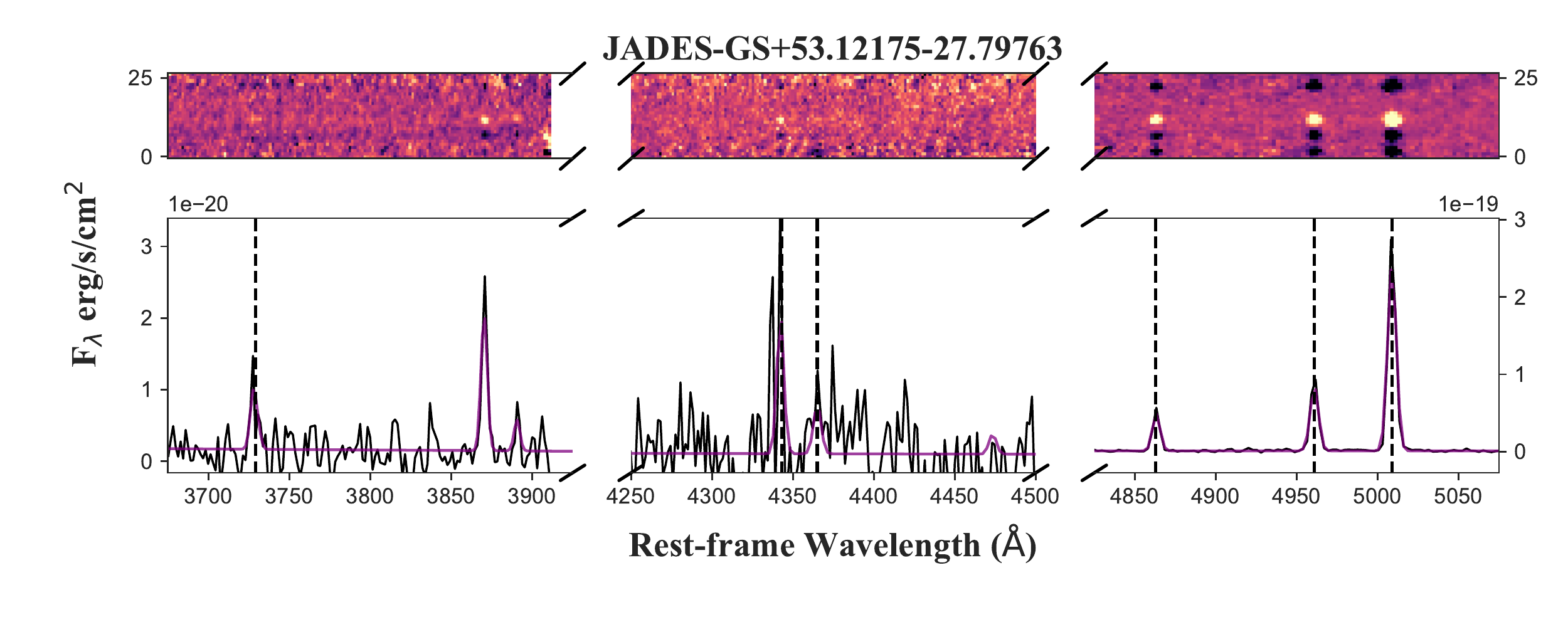}
    \includegraphics[width = \columnwidth]{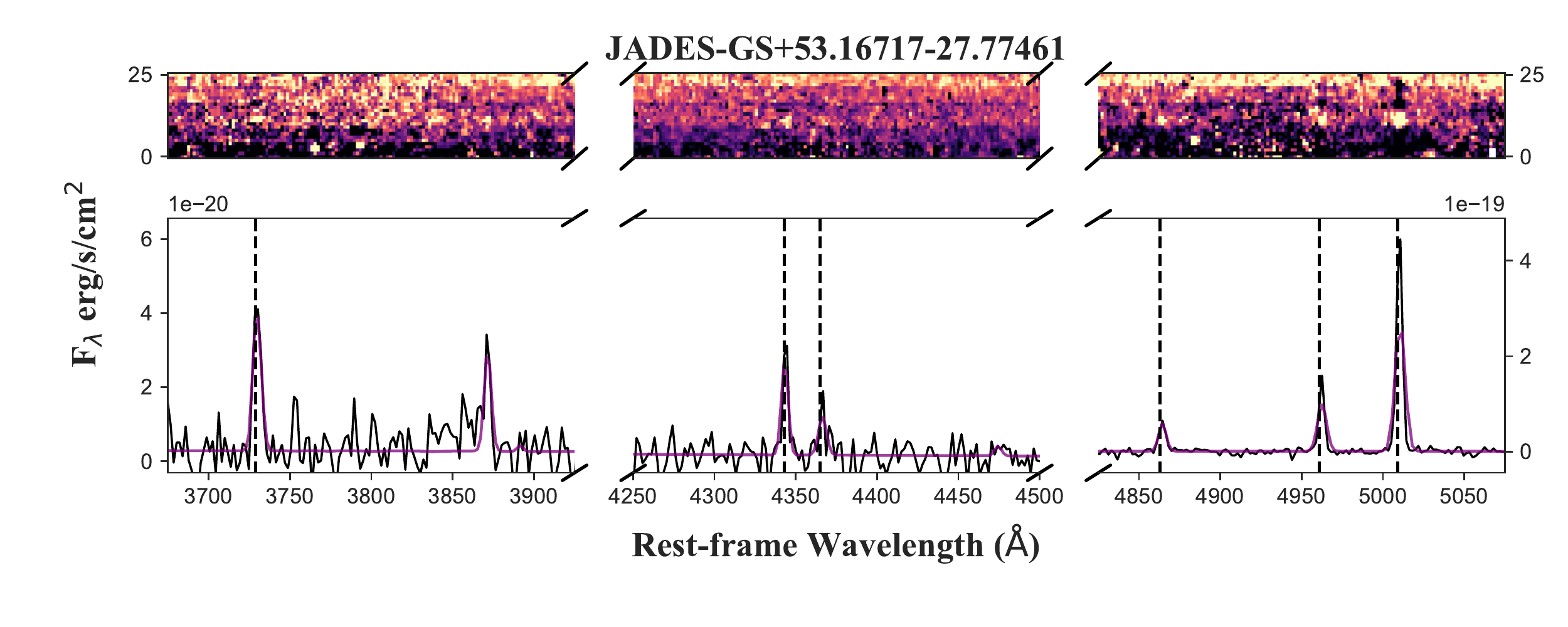}
    \includegraphics[width = \columnwidth]{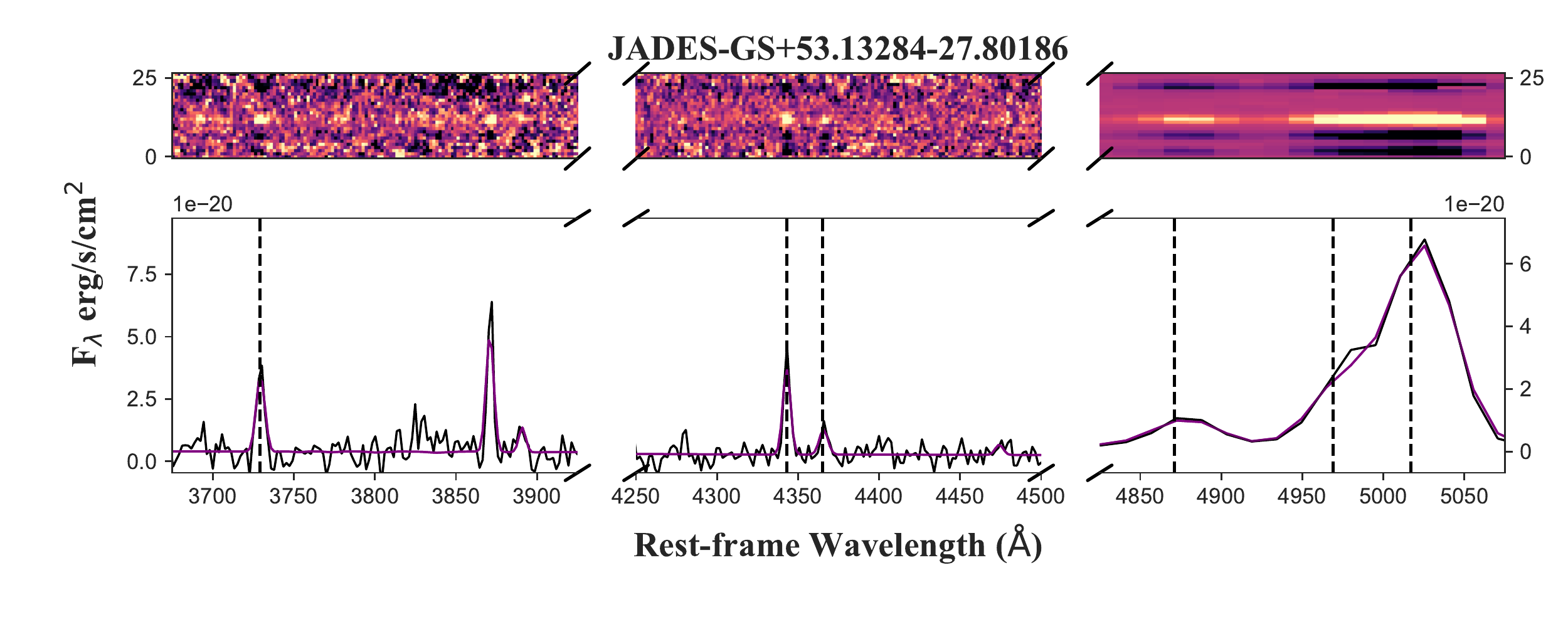}
    \includegraphics[width = \columnwidth]{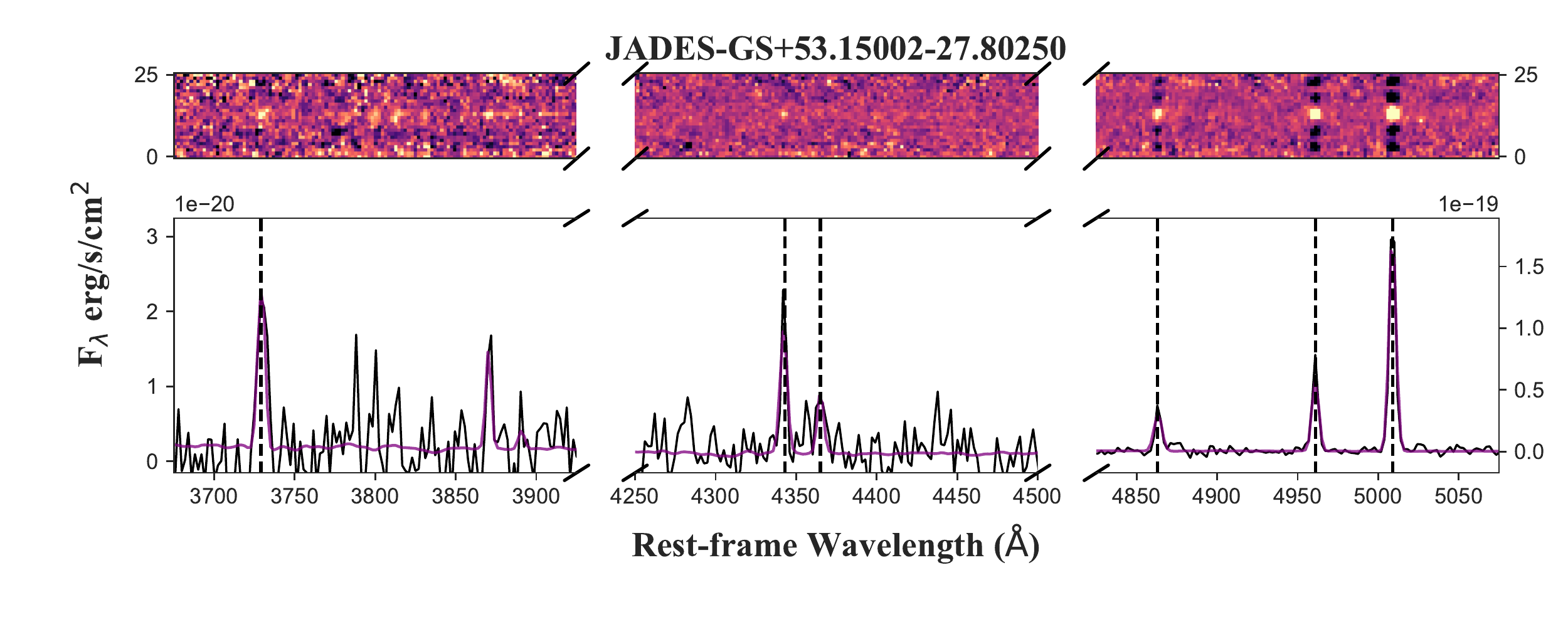}
    \includegraphics[width = \columnwidth]{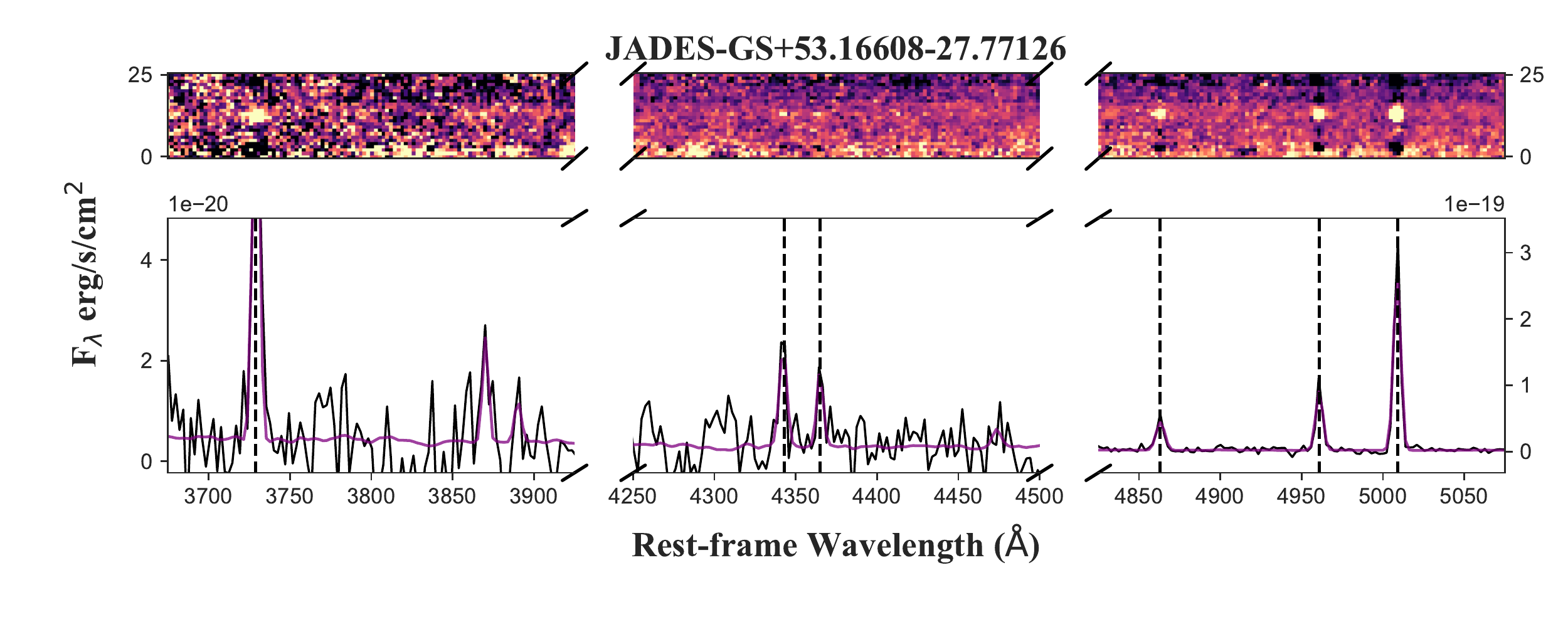}
    \includegraphics[width = \columnwidth]{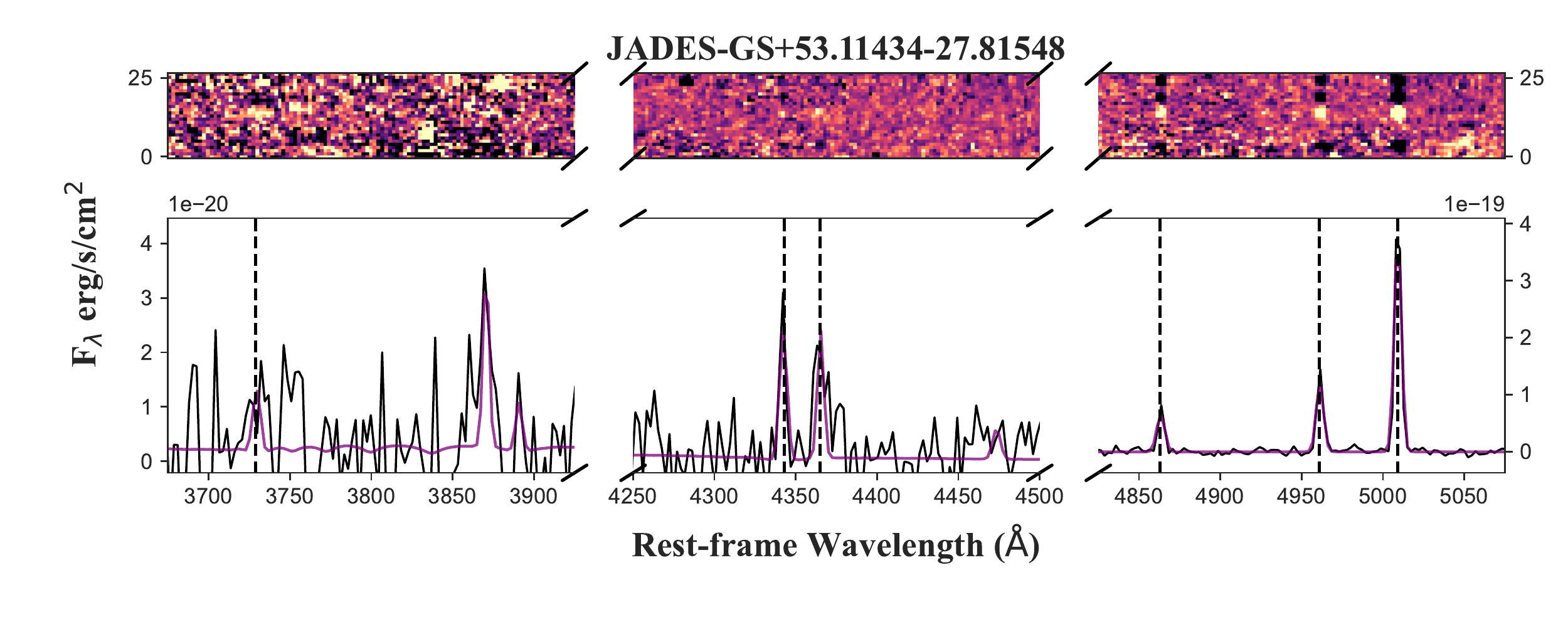}
    \includegraphics[width = \columnwidth]{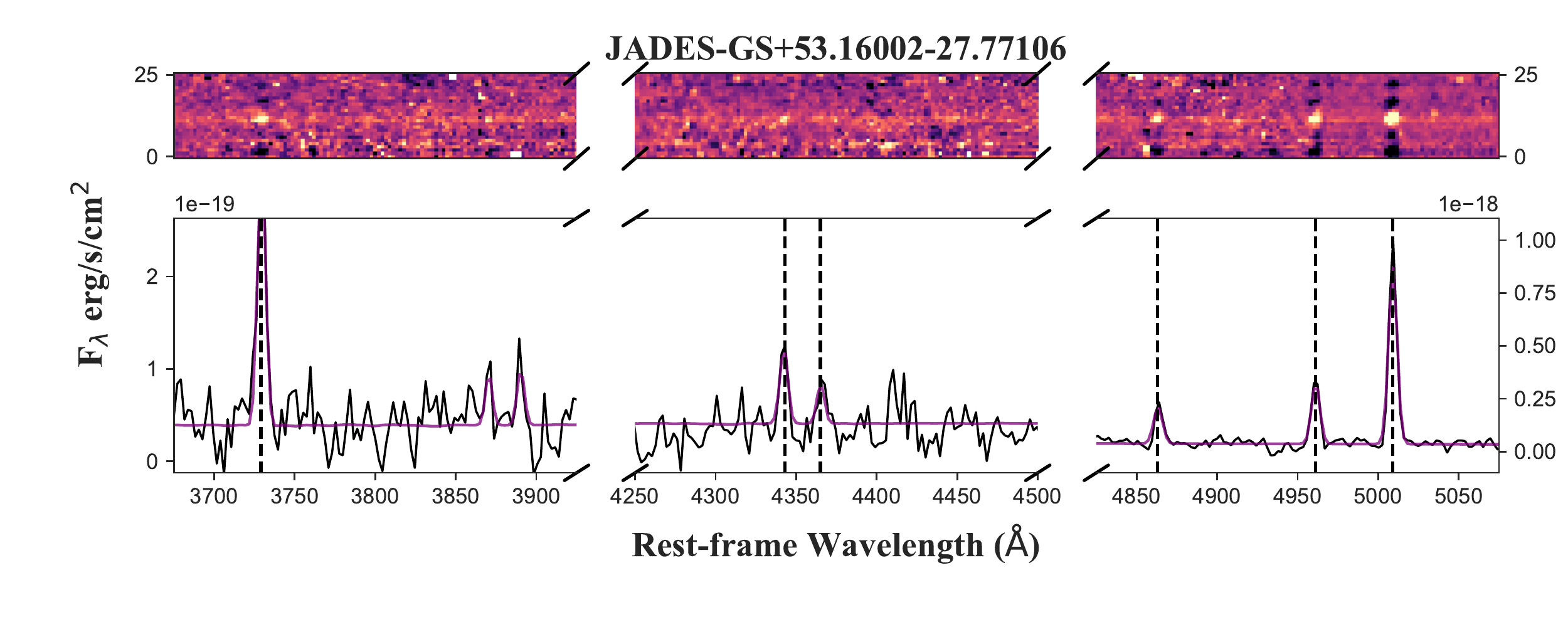} 
    \includegraphics[width = \columnwidth]{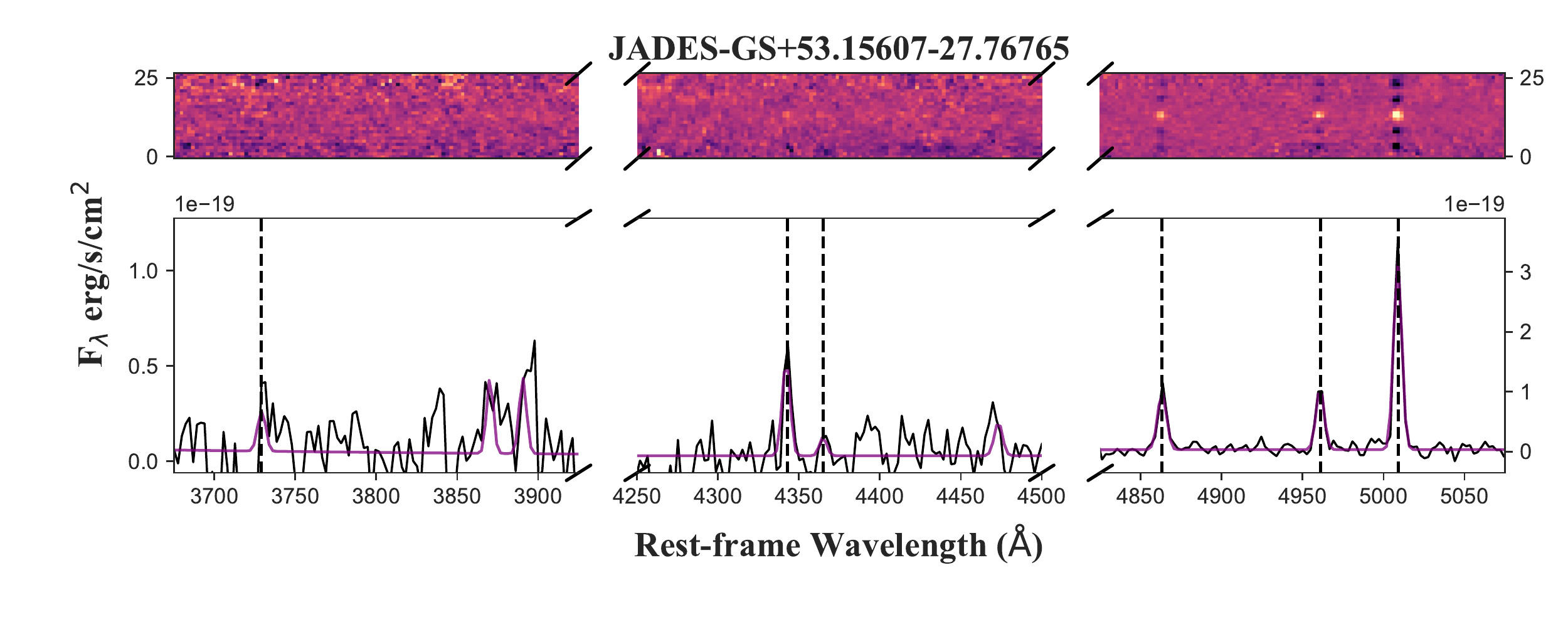}
    
    \caption{JWST/NIRSpec R1000 spectra of our 10 novel detections. The best-fit models from \texttt{PPXF} are shown in purple. The lines of interest (located at the vertical dotted lines) shown from left to right are [OII]$\lambda\lambda 3727, 3729$, H$\gamma$ $\&$ [OIII]$\lambda 4363$, and H$\beta$ $\&$ [OIII]$\lambda\lambda 4959, 5007$. [OII]$\lambda\lambda 3727, 3729$ and H$\gamma$ $\&$ [OIII]$\lambda 4363$ share the same y-axis. The top panels show the 2D spectrum for each respective line complex. JADES-GS+53.13284-27.80185 did not have [OIII]$\lambda\lambda 4959, 5007$ coverage in R1000 due to the detector gap, so we present [OIII]$\lambda\lambda 4959, 5007$ from our PRISM observations. }
    \label{fig:Spectra 1}
\end{figure*}

We visually inspect the 1D and 2D PRISM/CLEAR and grating spectra for our 253 unique targets and find 10 sources with an [OIII]$\lambda4363$ detection detected at a S/N $\gtrsim 3$. The median S/N in [OIII]$\lambda4363$ of our JADES sample is $\sim 5$. We present in Figure \ref{fig: Histogram} the redshift distribution of our parent sample and identified [OIII]$\lambda4363$ emitters. We show in Figure \ref{fig:Spectra 1} the [OII]$\lambda\lambda 3727, 3729$, $\text{H}\gamma$ and [OIII]$\lambda 4363$, and H$\beta$, [OIII]$\lambda\lambda 4959, 5007$ complexes of our [OIII]$\lambda4363$ sources. Object JADES-GS+53.13284-27.80186 has one of the highest S/N [OIII]$\lambda4363$ detection in our sample with a S/N $= 9.8$. However, [OIII]$\lambda\lambda4959, 5007$ fell within the detector gap for this object, so we instead use the [OIII]$\lambda\lambda4959, 5007$ fluxes from our PRISM observations. We correct for reddening in our measurements from the available Balmer lines adopting a \cite{Calzetti_2000} attenuation curve. We assume the theoretical ratios of H$\alpha / \text{H}\beta = 2.86$ and H$\beta / \text{H}\gamma = 2.13$ from Case B recombination at T$=1.5\times 10^4$K. We default to correcting with respect to H$\alpha / \text{H}\beta$, but we use H$\beta / \text{H}\gamma $ when H$\alpha$ is not available.

We can now determine electron temperatures and oxygen abundances through $T_e$. However, we note it is customary to take oxygen abundances as representative of the total gas-phase metallicity, which has implicit assumptions that all other chemical elements scale proportionally and that individual galaxies are a single HII region comprised of a high-ionization zone traced by O$^{++}$ and a low-ionization zone traced by O$^{+}$, which ignores the underlying temperature distribution and ionization structure. A detailed discussion of the nuance of these assumptions is outside the scope of this work (see \cite{Stasinska_2002} and \cite{Maiolino_2019} for a review), but there is novel work testing the significance of these assumptions \citep[e.g.,][]{Cameron_2022} that we are expanding upon.

Nonetheless, we derive the electron temperature for O$^{++}$ (t$_3$) by taking flux ratio of the [OIII]$\lambda\lambda4959, 5007$ doublet to the [OIII]$\lambda4363$ thermal line. We used \texttt{Pyneb} \citep{Luridiana_2015} with O$2+$ and O$+$ collision strengths from \cite{AK_1999} \& \cite{Palay_2012}  and \cite{Pradhan_2006} \& \cite{Tayal_2007}, respectively. A more problematic step is determining the electron temperature for O$^{+}$ (t$_2$). Only t$_3$ is derived directly here as we do not have spectral coverage of [OII] auroral lines at $7320$~\AA~and $7330$~\AA. In situations where [OII] auroral lines are not detected, it is common to interconvert between t3 and t2 using modeled relations. One such t$_3$ - t$_2$ relation is presented by \cite{Curti_2017} (originally presented in \cite{Pilyugin_2009}), in which they relate directly derived t$_3$ and t$_2$ temperatures to obtain the relation: t$_2 = 0.264 + t_3\times(0.835)$.

However, t$_3$ - t$_2$ relations have not been explored in the high-\textit{z} Universe. \cite{Yates_2020} found local t$_3$ - t$_2$ relations have difficultly in matching large samples of local galaxies with $T_e$ derived metallicities. Fortunately, there is typically little change in the total derived metallicity when adding O+ to O2+ as O2+ dominates the ionization state of oxygen in galaxies with direct [OIII]$\lambda4363$ detections \citep{Izotov_2006, Andrews_2013, Curti_2017, Curti_2020, Laseter_2022, Curti_2023}. Nonetheless, there is a clear need for future investigation of t$_3$ - t$_2$ relations in the high-\textit{z} Universe. 

\begin{table*}
\centering
    \resizebox{\textwidth}{!}{\begin{tabular}{ccccccc}
    \toprule
     JADES ID &  [OII]$\lambda 3727,29$ &  H$\gamma$ & [OIII]$\lambda 4363$ &  H$\beta$ & [OIII]$\lambda 4959$ &   [OIII]$\lambda 5007$\\
    \midrule
    JADES-GS+53.11243-27.77461 & 2.24 $\pm$ 0.63 & 9.42 $\pm$ 1.06 & 5.78 $\pm$ 1.12 & 25.29 $\pm$ 1.45 & 30.61 $\pm$ 1.31 & 92.23 $\pm$ 2.28\\
    ERO 4590 & 11.19 $\pm$ 4.66 & 45.58 $\pm$ 3.45 & 17.28 $\pm$ 3.31 & 134.23 $\pm$ 4.67 & 154.04 $\pm$ 5.21 & 459.82 $\pm$ 6.78\\
    ERO 6355 & 213.47 $\pm$ 6.38 & 108.49 $\pm$ 3.58 & 26.34 $\pm$ 3.08 & 238.72 $\pm$ 4.27 & 615.7 $\pm$5.81 & 1837.91 $\pm$ 8.5\\
    ERO 10612 & 33.5 $\pm$ 5.27 & 67.66 $\pm$ 3.2 & 21.98 $\pm$ 3.04 & 137.56 $\pm$ 3.79 & 307.63 $\pm$ 4.35 & 918.3 $\pm$ 6.63\\
    JADES-GS+53.13492-27.77271 & 33.64 $\pm$ 3.27 & 47.27 $\pm$ 3.58 & 17.39 $\pm$ 3.16 & 97.98 $\pm$ 3.47 & 192.81 $\pm$ 3.38 & 574.78 $\pm$ 5.82\\
    JADES-GS+53.12175-27.79763 & 10.40 $\pm$ 2.53 & 35.79 $\pm$ 1.43 & 14.95 $\pm$ 1.47 & 90.96 $\pm$ 1.62 & 165.64 $\pm$ 1.66 & 497.98 $\pm$ 2.89\\
    JADES-GS+53.16718-27.77462 & 995.65 $\pm$ 121.91 & 543.93 $\pm$ 38.28 & 156.79 $\pm$ 23.56 & 796.33 $\pm$ 38.22 & 1839.14 $\pm$ 38.09 & 5404.32 $\pm$ 64.79\\
    JADES-GS+53.13284-27.80185 & 6.18 $\pm$ 0.96 & 9.13 $\pm$ 0.34 & 3.38 $\pm$ 0.34 & 31.85 $\pm$ 0.66 & 72.93 $\pm$ 0.45 & 222.14 $\pm$ 0.78\\
    JADES-GS+53.15003-27.80251 & 27.10 $\pm$ 2.25 & 22.07 $\pm$ 2.27 & 13.39 $\pm$ 4.56 & 53.20 $\pm$ 1.83 & 91.79 $\pm$ 1.68 & 275.35 $\pm$ 2.91\\
    JADES-GS+53.16609-27.77126 & 32.78 $\pm$ 3.12 & 13.89 $\pm$ 1.67 & 11.42 $\pm$ 2.28 & 38.80 $\pm$ 1.10 & 69.48 $\pm$ 0.98 & 209.70 $\pm$ 1.72\\
    JADES-GS+53.11434-27.81549 & 0.31 $\pm$ 0.07 & 0.79 $\pm$ 0.12 & 0.92 $\pm$ 0.19 & 3.84 $\pm$ 0.15 & 7.56 $\pm$ 0.15 & 23.34 $\pm$ 0.27\\
    JADES-GS+53.16002-27.77107 & 1.59 $\pm$ 0.05 & 0.70 $\pm$ 0.06 & 0.35 $\pm$ 0.05 & 4.13 $\pm$ 0.16 & 6.10 $\pm$ 0.13 & 19.00 $\pm$ 0.24\\
    JADES-GS+53.15608-27.76766 & 81.96 $\pm$ 8.22 & 108.91 $\pm$ 6.92 & 21.89 $\pm$ 6.06 & 214.04 $\pm$ 8.53 & 222.43 $\pm$ 6.14 & 661.64 $\pm$ 10.57\\

    \bottomrule
    \end{tabular}}
    \caption{Measured fluxes and errors of emission lines of interest from \texttt{PPXF} in units of 10$^{-20}$ erg/s/cm$^2$.}
    \label{Fluxes Table}
\end{table*}

We determine ionic oxygen abundances using \texttt{Pyneb} with the same collision strengths as before. We assume an electron density of $N_e = 300 \text{cm}^{-3}$ since this is representative of the ISM electron density of \textit{z} $\sim$ 2-3 galaxies \citep{Sanders_2016a, Sanders_2016b}. The choice of electron density does not significantly affect the temperature results. For example, when assuming $N_e = 1,000 \text{cm}^{-3}$, there is $\sim 0.1\%$ change in the derived t$_3$ \citep{Izotov_2006}. We determine the total oxygen abundance for each galaxy by taking ($\frac{\text{O}}{\text{H}} = \frac{\text{O+}}{\text{H}} + \frac{\text{O2+}}{\text{H}}$). We do not detect any HeII $\lambda 4686$ in our sample, so we do not apply an ionization correction factor to account for O3+ since HeII has an ionization potential of $\gtrsim 54.4$ eV and O3+ has an ionization potential of $\gtrsim 55$ eV. Even if O3+ is present, a correction would have nominal change for the total oxygen abundance \citep{Izotov_2006, Berg_2021, Curti_2023}. 

To calculate the uncertainties of our measurements we use a Monte Carlo technique. We evaluate the electron temperature and oxygen abundance 10,000 times using values drawn randomly from normal distributions for the measured fluxes of [OIII]$\lambda\lambda\lambda 5007,4959,4363$, H$\beta$, H$\gamma$, and [OII]$\lambda\lambda 3727, 3729$, centered at the measured flux values, and with standard deviations corresponding to the $1\sigma$ flux errors from \texttt{PPXF}. Our final reported electron temperatures and metallicities are taken as the median value of the propagated normal distributions with the standard deviation of the distributions being the $1\sigma$ error. 

In addition to our [OIII]$\lambda 4363$ emitters, \cite{Curti_2023} measured the chemical abundances of three $z \sim 8$ galaxies behind the galaxy cluster SMACS J0723.3-7327 during the initial ERO data release. A number of studies investigated the same objects \citep[e.g.,][]{Schaerer_2022, Taylor_2022, Rhoads_2023, Trump_2023}. However, \cite{Curti_2023} reprocessed the data through the NIRSPec GTO pipeline. We include these three galaxies (ID: 4590, 6355, and 10612) after reprocessing the initial data from \cite{Curti_2023} with the updated NIRSpec GTO pipeline (Carniani et al., in preparation) and determining oxygen abundances as described above. We find nominal changes in the total metallicities: $0.24$~dex for 4590, $-0.1$~dex for 6355, and $0.04$~dex for 10612. For our combined sample we report the line fluxes in Table \ref{Fluxes Table} and electron temperatures/metallicites in Table \ref{Table: Galaxy Properties}.

Recently, \cite{Bunker_2023} provided the first JWST/NIRSpec spectrum of GN-z11 \citep{Oesch_2016} from the JADES collaboration. \cite{Bunker_2023} reports a detection of [OIII]$\lambda 4363$, but there was insufficient wavelength coverage to observe [OIII]$\lambda \lambda 4959, 5007$, thus we cannot use the $T_e$ method. The proceeding analysis and subsequent discussion in Sections \ref{Strong Line Calibrations} and \ref{Discussion} require a self-consistent metallicity prescription. Therefore, we do not include GN-z11 in our sample, but we highlight the detection of [OIII]$\lambda 4363$ in the most luminous Lyman break galaxy at $z > 10$ for context in our discussion in Section \ref{Discussion}.

\begin{table*}
    \centering
    \begin{tabular}{cccccc}
    \toprule
     JADES ID & \textit{z} & $T_e (10^4 \text{K})$  & $12+\log(\text{O/H})$ & EW$_0$(H$\beta$) & S/N ([OIII]$\lambda 4363$) \\
    \midrule
    JADES-GS+53.11243-27.77461 & 9.43 & 3.16 $\pm$ 0.69 & 7.03 $\pm$ 0.10 &  85.53 $\pm $ 4.91 & 5.1 \\
    ERO 4590  &  8.496  & 2.15 $\pm$ 0.28 & 7.23 $\pm$ 0.11 & 107.89 $\substack{+74.01 \\ -36.73}$ $^{\dagger}$ & 5.2\\
    ERO 6355 &  7.665  & 1.32 $\pm$ 0.06 & 8.14 $\pm$ 0.06 & 174.00 $\substack{+86.03 \\ -6.88}$ $^{\dagger}$ & 8.5\\
    ERO 10612 &  7.658 & 1.65 $\pm$ 0.01 & 7.77 $\pm$ 0.07 & 351.37  $\substack{+515.90 \\ -111.61}$ $^{\dagger}$ & 7.2\\
    JADES-GS+53.13492-27.77271 & 6.33 & 1.82 $\pm$ 0.20 & 7.58 $\pm$ 0.12 & 83.69 $\pm $ 2.96 & 5.5\\
    JADES-GS+53.12175-27.79763 & 5.94 & 1.81 $\pm$ 0.11 & 7.54 $\pm$ 0.06 & 71.04 $\pm $ 1.27 & 10.2\\
    %JADES-GS+53.12972-27.80818 & 5.56 & 4.05 $\pm$ 2.53 &  7.37 $\pm$ 0.19 &  23.61 $\pm $ 3.50 & 1.5$^{\dagger}$\\
    JADES-GS+53.16718-27.77462 & 4.77 & 1.77 $\pm$ 0.16 & 7.73 $\pm$ 0.10 & 239.49 $\pm$ 11.49 & 6.7\\
    JADES-GS+53.13284-27.80186 & 4.65 & 1.29 $\pm$ 0.05 & 8.04 $\pm$ 0.06 & 301.88 $\pm$ 6.23 & 9.8\\
    JADES-GS+53.15002-27.80250 & 4.23 & 2.51 $\pm$ 0.76 & 7.34 $\pm$ 0.23 & 122.53 $\pm $ 4.20 & 2.9$^\ddag$\\
    JADES-GS+53.16609-27.77126 & 3.60 & 2.80 $\pm$ 0.50 & 7.28 $\pm$ 0.12 &  77.19 $\pm $  2.18 & 5.0\\
    JADES-GS+53.11434-27.81549 & 3.59 & 2.15 $\pm $ 0.31 & 7.41 $\pm $ 0.13 & 250.93 $\pm $ 10.12 & 4.9\\
    JADES-GS+53.16002-27.77107 & 1.85 & 1.40 $\pm$ 0.10 & 7.78 $\pm$ 0.09 &  222.64 $\pm $ 8.78 & 6.4\\
    JADES-GS+53.15608-27.76766 & 1.72 &	1.92 $\pm$	0.33 &	7.27 $\pm$ 0.20 & 552.35 $\pm$ 22.00 & 3.6\\
    
    \bottomrule
    \end{tabular}
    \caption{Derived galaxy properties of our sample. $\dagger$ denotes EW$_0$(H$\beta$) values taken from \cite{Taylor_2022}. $\ddag$ S/N is $< 3$, but we find clear detection of [OIII]$\lambda 4363$, so we include JADES-GS+53.15002-27.80250 in our sample. }
    \label{Table: Galaxy Properties}
\end{table*}

\subsection{\textbf{CEERS}}

\subsubsection{\textbf{Comparison}} \label{Pipeline Comparison}

Recently, \cite{Sanders_2023} identified [OIII]$\lambda 4363$ in 16 galaxies between $z \approx 2.0-9.0$, measured from JWST/NIRSpec observations obtained as part of the Cosmic Evolution Early Release Science (CEERS) survey program. They further consolidated 9 objects with [OIII]$\lambda 4363$ detections between $z \approx 4-9$ from the literature using JWST/NIRSpec along with 21 galaxies between $z \approx 1.4-3.7$ with detections from ground-based spectroscopy.  \cite{Sanders_2023} determined metallicities with $T_e$ through \texttt{PyNeb} \citep{Luridiana_2015} for their entire sample to construct empirical $T_e$-based metallicity calibrations for strong-line ratios such as R2, O3O2, R3, and R23 in the high-\textit{z} Universe, which we investigate in Section \ref{Strong Line Calibrations}. As such, we include the 16 discovered galaxies with  [OIII]$\lambda 4363$ from CEERS in our comparisons. However, \cite{Sanders_2023} used O2+ and O+ collision strengths from \cite{Storey_2014} and \cite{Kisielius_2009}, respectively. We re-derive the metallicities for the \cite{Sanders_2023} sample using O2+ and O+ collision strengths from \cite{AK_1999} \& \cite{Palay_2012} and \cite{Pradhan_2006} \& \cite{Tayal_2007} to remain self-consistent. We investigate the systematics of choosing different O2+ collisional strengths in the Appendix A. 

A caveat with including the sample from \cite{Sanders_2023} is the difference in spectroscopic reduction pipelines employed. Specifically, data were reduced in \cite{Sanders_2023} with \texttt{calwebb detector}, STScI's pipeline, whereas we utilize the GTO pipeline as mentioned in Section \ref{Data Processing}. Issues and variations between the pipelines were immediately apparent from the works of \cite{Schaerer_2022, Taylor_2022, Rhoads_2023, Trump_2023}; and \cite{Curti_2023}, with overall conclusions being that analyses and interpretations should avoid absolute flux calibrations and using widely separated line ratios \citep{Trump_2023}. Recently, \cite{Maseda_2023} provided deeper insight into these discrepancies. However, GTO flux calibrations have improved since these studies, though a full description will be presented in Bunker et al. (in preparation). A full comparison between the current strengths and weaknesses of the pipelines are outside the scope of this work, but for the current comparison between our JADES sample and \cite{Sanders_2023}, systematics could exacerbate or diminish offsets between metallicity determinations and strong-line ratios.  

\subsubsection{\textbf{Metallicity Prescription Choice}} \label{Metallicity Prescription Choice}

In addition to systematics introduced through data reduction and the choice in collisional strengths, the decision to use a given metallicity prescription will introduce systematics, amongst other choices (e.g., the t$_3$-t$_2$ relation). We demonstrate these systematics by re-deriving electron temperatures and metallicities for our JADES sample and the \cite{Sanders_2023} sample using the \cite{Izotov_2006} prescription. We use the atomic data listed in \cite{Stasinska_2005} to determine t$_3$ in an iterative manner (\cite{Izotov_2006} equations 1 and 2). We derive t$_2$ using equation 14 from \cite{Izotov_2006}, which was obtained by relating t$_3$ to temperatures of other ions from photoionization models that best fit HII emission line observations \citep{Izotov_2006}. 

\begin{figure}[hbt!]
    \centering
    \includegraphics[width = \columnwidth]{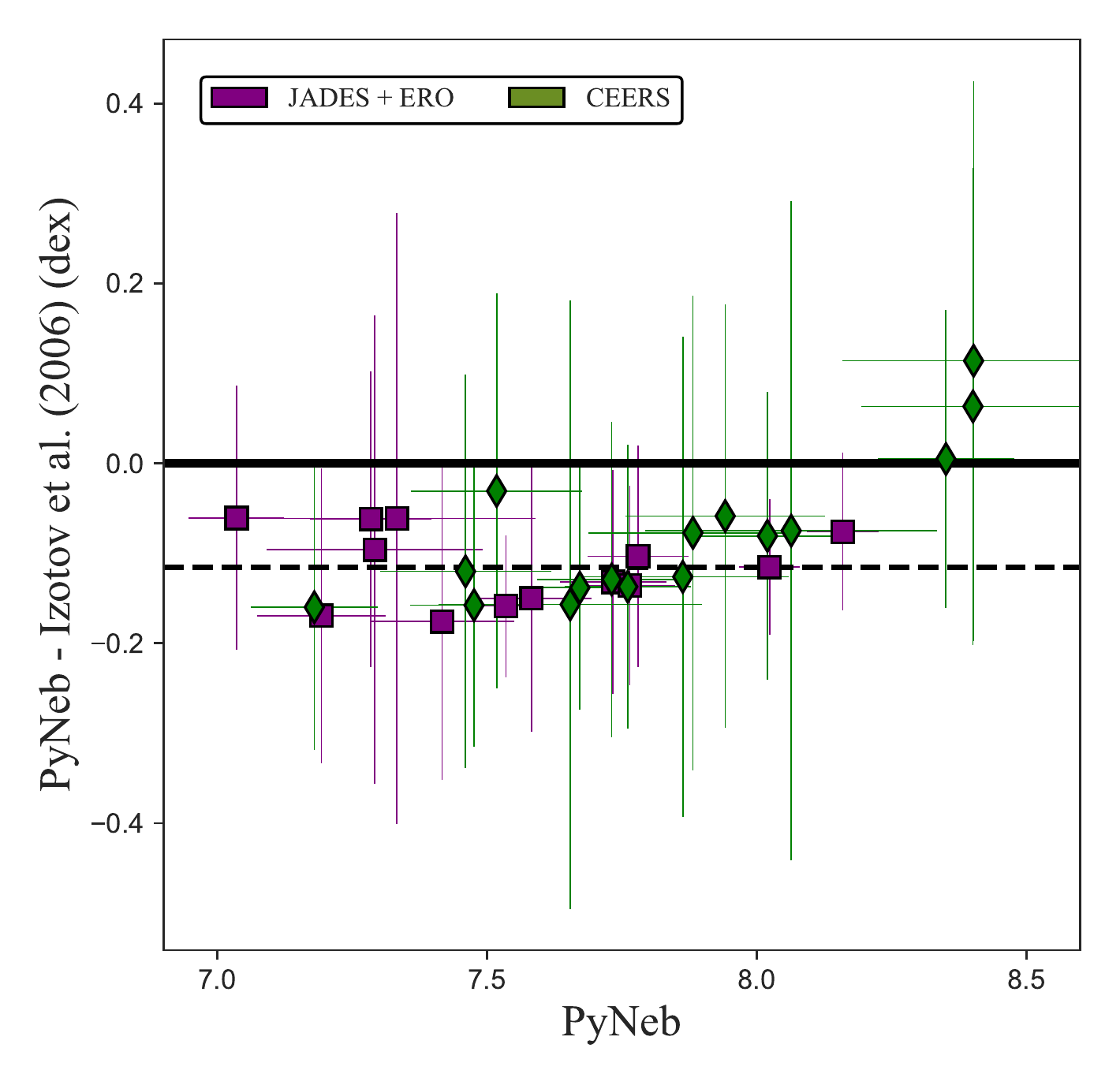}
    \caption{Deviation between metallicites derived by \cite{Izotov_2006} and \texttt{PyNeb}. The solid line represents unity, whereas the dashed line represents the median offset between \cite{Izotov_2006} and \texttt{PyNeb}.}
    \label{fig:Systematics}
\end{figure}

We present in Figure \ref{fig:Systematics} the systematic offsets between \cite{Izotov_2006} and \texttt{PyNeb} derived metallicities for our sample. We find a median offset of $\Delta 12 + \log(\text{O/H})$ of $-0.11$~dex when using \texttt{PyNeb} instead of \cite{Izotov_2006}. A critical assessment of the advantages and limitations of $T_e$ metallicity prescriptions is outside the scope of this work. However, it is clear that choice does matter, thus demonstrating the need for \textit{self-consistency} in metallicity studies and comparisons as [OIII]$\lambda 4363$ samples in the high-\textit{z} Universe continue to grow. We continue with the analysis using metallicities derived with \texttt{Pyneb}. We include in Appendix A the Figures presented in Section \ref{Strong Line Calibrations} for \cite{Izotov_2006} derived metallicities. Nonetheless, the main results discussed in Section \ref{Discussion} remain unchanged irregardless of the $T_e$ method employed. 

%%%%%%%%%%%%%%%%%%%%%%%%%%%%%%%%%%%%%%%%%%%%%%%%%%%%%%%%%

\section{\textbf{Strong Line Calibrations}} \label{Strong Line Calibrations}

\subsection{\textbf{Comparison to Locally Derived Strong Line Calibrations}} \label{Comparison to Locally Derived Strong Line Calibrations}

As mentioned in Section \ref{Introduction}, there are a number of strong nebular emission-line ratios calibrated against T$_e$ derived metallicities to act as metallicity diagnostics \citep[e.g.,][]{Pettini_2004, Maiolino_2008, Marino_2013, Pilyugin_2016, Curti_2017, Bian_2018, Sanders_2021}. These calibrations have been applied on large samples of galaxies to determine metallicities when auroral lines are not observed, which allows for larger characteristic studies, such as the MZR \citep[e.g.,][]{Tremonti_2004, Mannucci_2010, Perez_2013, Lian_2015, Maiolino_2019, Curti_2020, Baker_2023} and the Fundamental Metallicity Relation (FMR) \citep[e.g.,][]{Mannucci_2010, Lara_2010, Brisbin_2012, Hunt_2012, Yates_2012, Nakajima_2014, Baker_2023b}. All calibrations have caveats, however, such as high dependencies on ionization parameter \citep[e.g.,][]{Dopita_2006, Pilyugin_2016} or an inherent assumption on the N/O–O/H relation \citep[e.g.,][]{Dopita_2016, Hayden_2022, Schaefer_2020, Schaefer_2022}. Another major uncertainty is the applicability of these strong line calibrations for high-\textit{z} galaxies. An evolution in the ISM conditions of high-redshift galaxies compared to the local Universe might impact the intrinsic dependence of strong-line ratios on gas-phase metallicity, potentially hampering their use as abundance diagnostics at high redshift, and thus biasing the assessment and interpretation of the chemical evolution history of galaxies. 

Already, \cite{Cameron_2023}, using the same parent data set as the current work, found $z \sim 5.5 - 9.5$ galaxy emission line ratios are generally consistent with galaxies with extremely high ionization parameters (log(U) = -1.5) and are traced by the extreme ends of $z \sim 0$ ionization-excitation diagrams of R23-O3O2 and R23-Ne3O2. In addition, \cite{Cameron_2023} found more than an order of magnitude of scatter in line ratios such as [OII]$\lambda\lambda 3727,3729$/H$\beta$ and [OIII]$\lambda 5007$/[OII]$\lambda\lambda 3727,3729$ while simultaneously not observing any [NII]$\lambda 6583$, indicating significant diversity in metallicity and ionization within the ISM conditions of the sample. To complicate the landscape, recent JADES/NIRSpec observations of the GN-z11, which is also an [OIII]$\lambda 4363$ emitter, revealed rarely-seen NIV]$\lambda 1486$ and NIII]$\lambda 1748$ lines that may imply an unusually high N/O abundance \citep{Bunker_2023, Cameron_2023_gnz11, Senchyna_2023}. 

Here, we utilize the T$_e$ derived abundances and emission line ratios delivered by the ‘Deep’ spectroscopic tier of JADES to provide a more detailed look at strong line calibrations in the high-\textit{z} Universe. We include the aforementioned ERO objects from \cite{Curti_2023} and the CEERS objects from \cite{Sanders_2023} derived in a self-consistent manner for a complete JADES+ERO+CEERS data set. We investigate some of the most widely adopted strong-line metallicity diagnostics:

\begin{gather*}
    \text{R}2~= \log (\frac{[\text{OII}]\lambda\lambda 3727,3729}{\text{H}\beta}),\\
    \text{O}3\text{O}2~= \log (\frac{[\text{OIII}]\lambda 5007}{[\text{OII}]\lambda\lambda 3727,3729}),\\
    \text{R}3~= \log (\frac{[\text{OIII}]\lambda 5007}{\text{H}\beta}),\\
    \text{R}23~= \log (\frac{[\text{OII}]\lambda\lambda 3727,3729 + [\text{OIII}]\lambda\lambda 4959, 5007}{\text{H}\beta}).\\
\end{gather*}
A common strong-line calibration, especially at high-\textit{z}, is N2 = [NII]$\lambda 6583$/H$\alpha$. We exclude this diagnostic from this study, however, because we find no  convincing evidence for [NII]$\lambda 6583$, analogous to \cite{Cameron_2023}.

\begin{figure*}[hbt!]
    \centering
    \includegraphics[width = \textwidth]{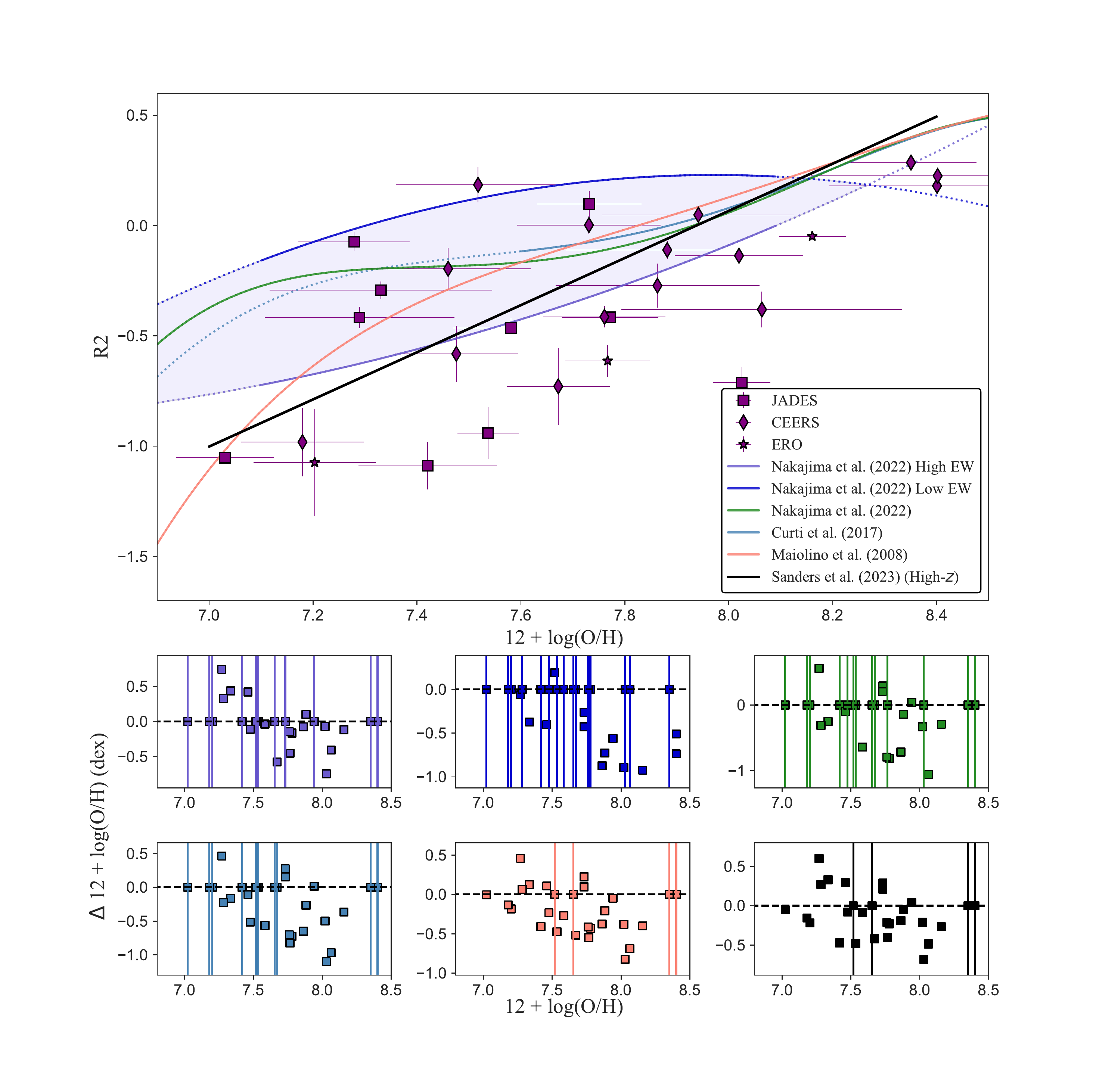}
    \caption{The relationship between $T_e$ metallicity and R2 for our JADES sample compared with strong-line calibrations from \cite{Maiolino_2008}, \cite{Curti_2017, Curti_2020}, and the ``All", ``Large Equivalent Width (EW)", and ``Small EW" calibrations from \cite{Nakajima_2022}. \cite{Bian_2018} does not include a calibration for R2, but we include their calibrations for O3O2, R3, and R23 in Figures \ref{fig:O3O2 Strong Line Comparison} - \ref{fig:R23 Strong Line Comparison}. Solid lines indicate calibrated ranges whereas dotted lines indicate the extrapolation of the calibration over the metallicity range $6.9 \leq 12 + \log(\text{O/H}) \leq 9.0$. The six subplots demonstrate the change between $T_e$ derived metallicities and calibration derived metallicities for our individual galaxies. The vertical lines represent the failure of a strong-line calibration to account for the measured line ratios at the given metallicity.}  
    \label{fig:R2 Strong Line Comparison}
\end{figure*}

\begin{figure*}[hbt!]
    \centering
    \includegraphics[width = \textwidth]{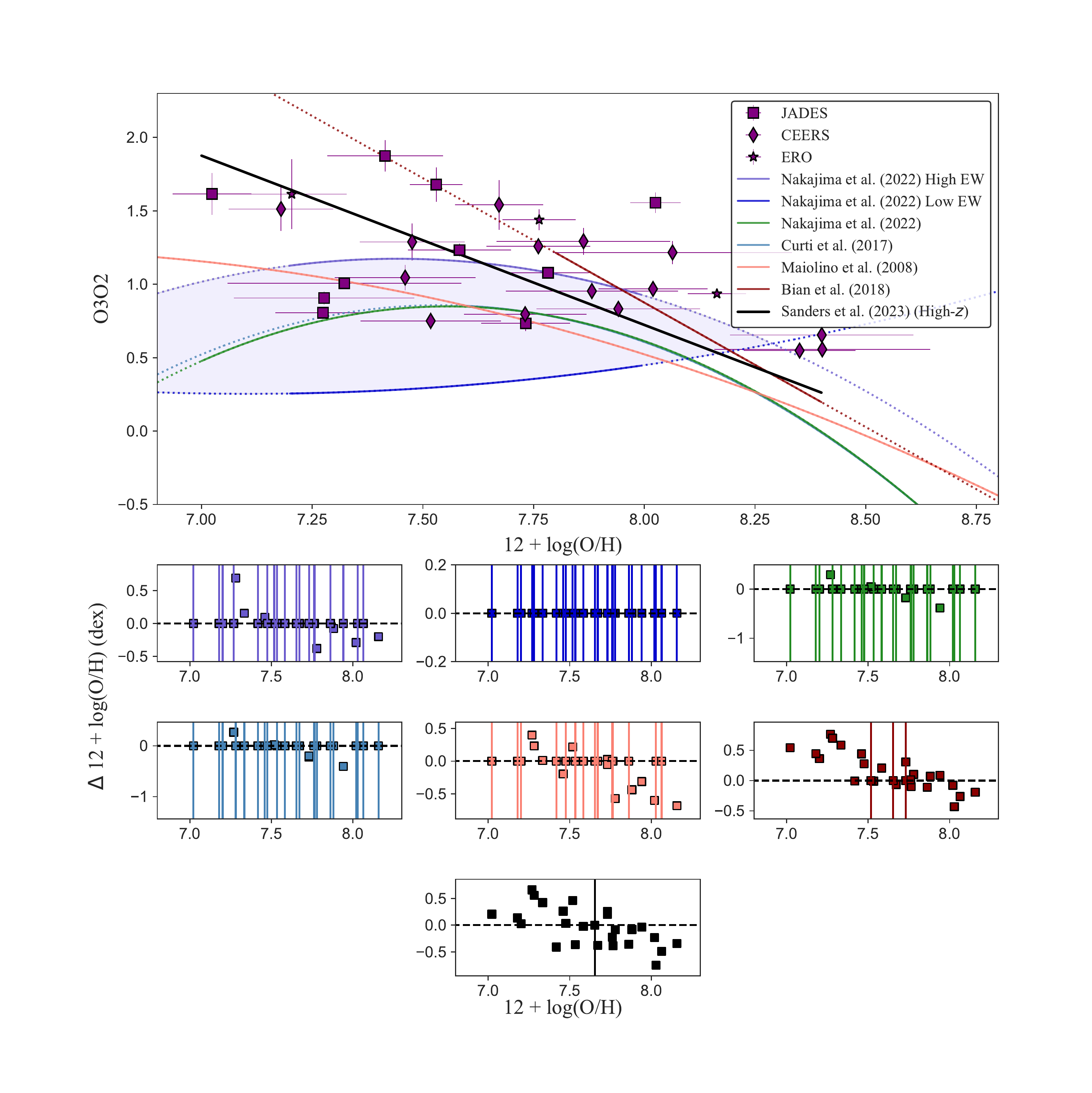}
    \caption{Identical to Figure \ref{fig:R2 Strong Line Comparison} except the relationship is between $T_e$ metallicity and O3O2.}
    \label{fig:O3O2 Strong Line Comparison}
\end{figure*}

\begin{table*}
    \centering
    \begin{tabular*}{\textwidth}{rrrrrrrrrrrrr}
    \multicolumn{13}{c}{\large Deviation of Local Calibrations from our JADES Sample (in units of $\sigma$)}\\
    \toprule
    \multicolumn{1}{c}{}  & \multicolumn{4}{c}{\cite{Maiolino_2008}} & \multicolumn{4}{c}{\cite{Curti_2017, Curti_2020}} & \multicolumn{4}{c}{\cite{Bian_2018}}\\
    \cmidrule(lr){2-5} \cmidrule(lr){6-9} \cmidrule(lr){10-13}
    &  R2 & R3 & R23 & O3O2 &  R2 & R3 & R23 & O3O2 &  R2 & R3 & R23 & O3O2\\
    $\sigma_{cal}$$^1$ &  0.10 & 0.10 & 0.06 & 0.20 &  0.11 & 0.09 & 0.06 & 0.15 &  $--$ & 0.15 & 0.15 & 0.15\\
    Sample Deviation$^2$ & 1.17 & 1.12 & 1.04 &  1.11 & 1.13 & 0.95 & 1.04 & 1.11 & $--$ & 0.50 & 0.58 & 0.96\\
    \midrule
    \midrule
    
    \multicolumn{1}{c}{}  & \multicolumn{4}{c}{\cite{Nakajima_2022} All} & \multicolumn{4}{c}{\cite{Nakajima_2022} Large EW} & \multicolumn{4}{c}{\cite{Nakajima_2022} Small EW}\\
    \cmidrule(lr){2-5} \cmidrule(lr){6-9} \cmidrule(lr){10-13}
    &  R2 & R3 & R23 & O3O2 &  R2 & R3 & R23 & O3O2 &  R2 & R3 & R23 & O3O2\\
    $\sigma_{cal}$ & 0.27 & 0.16 & 0.10 & 0.39 & 0.21 & 0.06 & 0.06 & 0.25 & 0.21 & 0.17 & 0.08 & 0.35 \\
    Sample Deviation & 1.05 & 1.05 & 1.04 & 1.13 & 1.09 & 0.66 & 0.74  & 0.96 & 1.32 & 1.67 & 1.28 & 1.60\\

    \bottomrule
    \end{tabular*}
    \caption{Significance of deviation (in units of $\sigma$) for the expected line ratios from each strong-line calibration to our JADES sample presented in Figures \ref{fig:R2 Strong Line Comparison} - \ref{fig:R23 Strong Line Comparison}. The metallicity dependency varies across each strong-line diagnostic examined (e.g., the turnover points in R3 and R23), thus we include $(12 + \log(\text{O/H})_{T_e}) - (12 + \log(\text{O/H})_{cal})$ in Figures \ref{fig:R2 Strong Line Comparison} - \ref{fig:R23 Strong Line Comparison} to demonstrate offsets with the respect to $12 + \log(\text{O/H})_{T_e}$ for our individual galaxies.}
    \label{Table: Deviation}
\end{table*}

\begin{figure*}[hbt!]
    \centering
    \includegraphics[width = \textwidth]{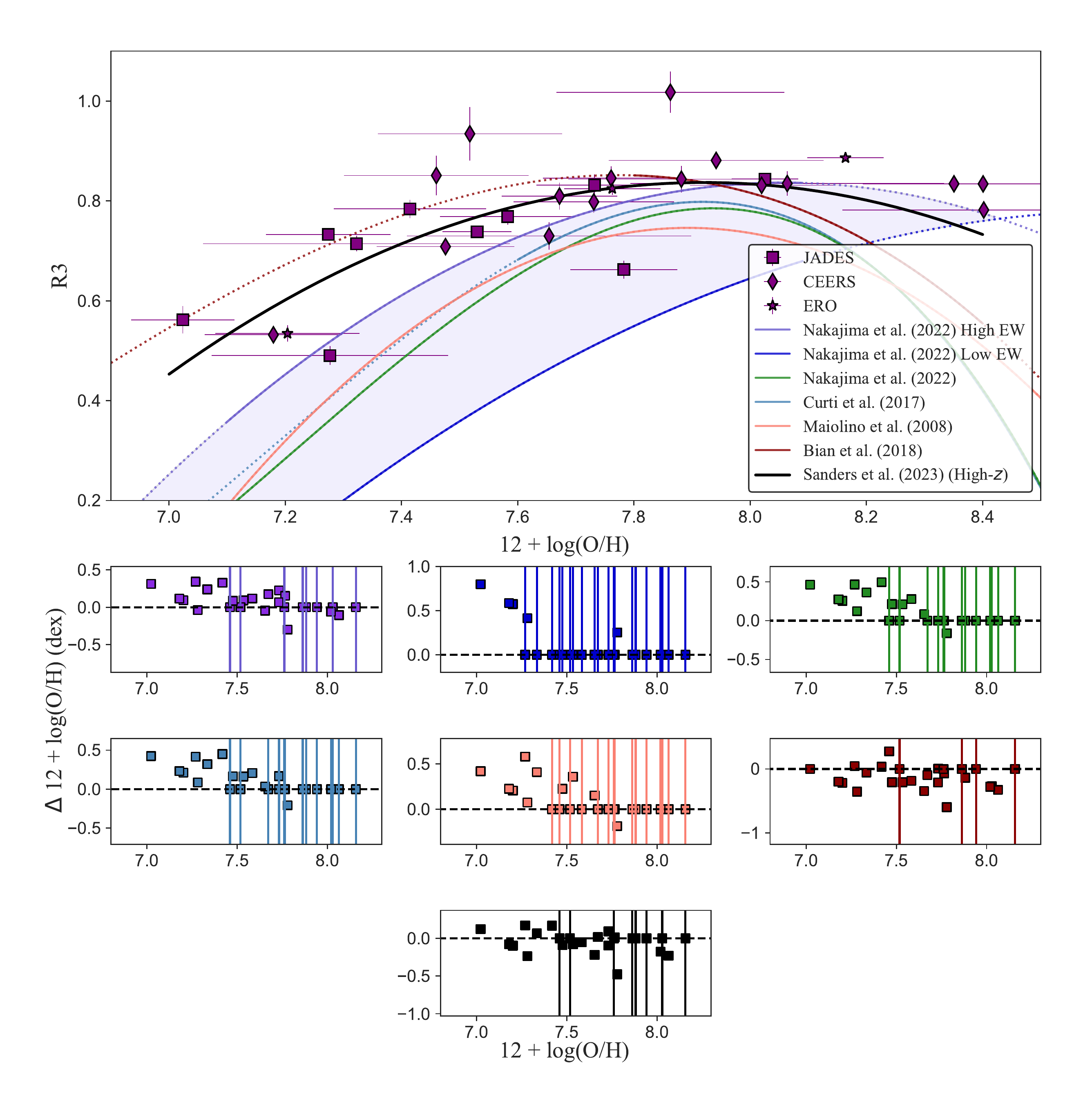}
    \caption{Identical to Figure \ref{fig:R2 Strong Line Comparison} except the relationship is between $T_e$ metallicity and R3.}
    \label{fig:R3 Strong Line Comparison}
\end{figure*}

\begin{figure*}[hbt!]
    \centering
    \includegraphics[width = \textwidth]{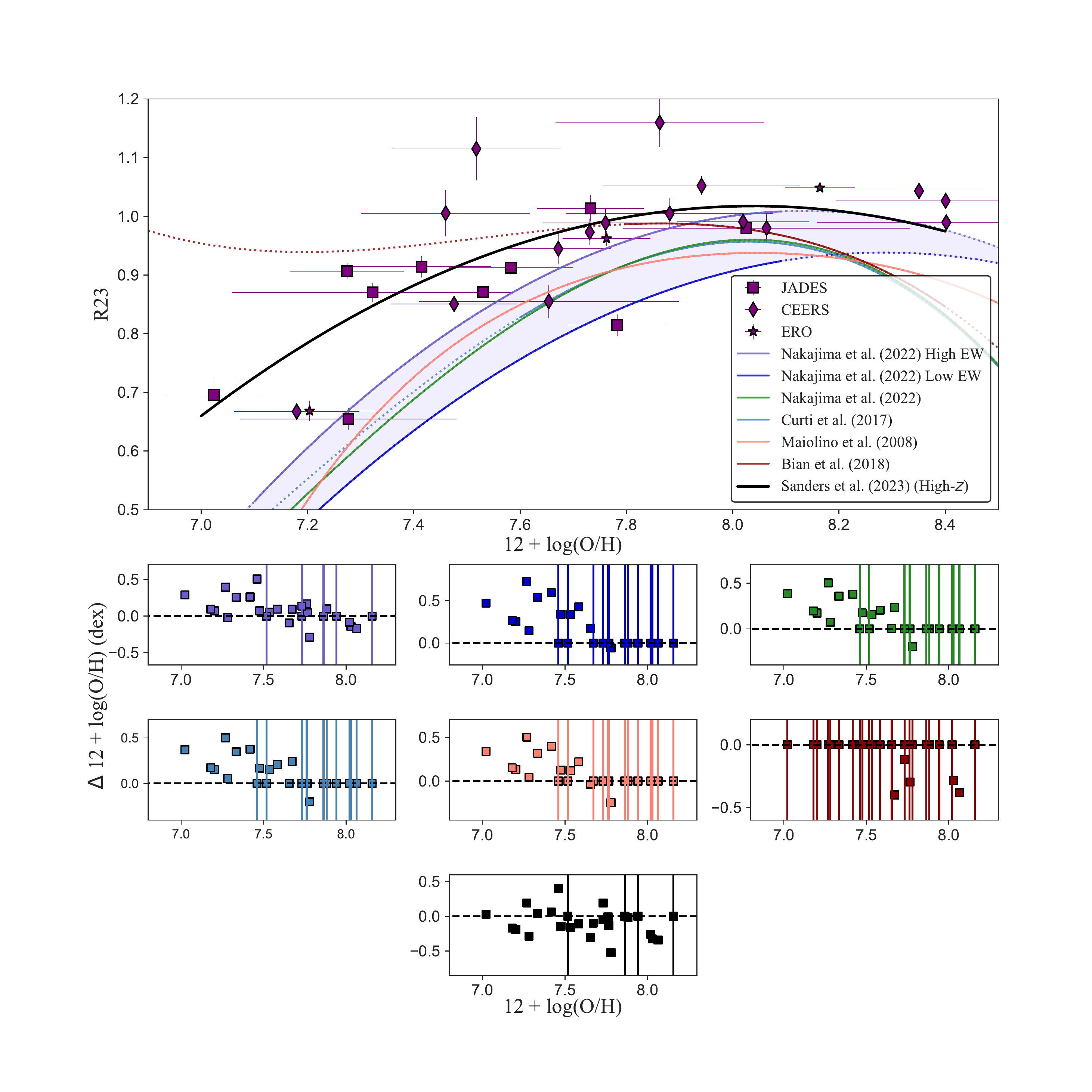}
    \caption{Identical to Figure \ref{fig:R2 Strong Line Comparison} except for the relationship between $T_e$ metallicity and R23.}
    \label{fig:R23 Strong Line Comparison}
\end{figure*}

We present in Figures \ref{fig:R2 Strong Line Comparison}-\ref{fig:R23 Strong Line Comparison} the strong-line ratios of our sample and the \cite{Sanders_2023} sample plotted against metallicity and an array of locally-derived strong-line calibrations. Specifically, we include \cite{Maiolino_2008}, \cite{Curti_2017, Curti_2020}, \cite{Bian_2018}, and \cite{Nakajima_2022}. In brief, \cite{Curti_2017, Curti_2020} provided calibrations based on T$_e$ metallicity measurements derived from SDSS stacked spectra and direct [OIII]$\lambda 4363$ detections. \cite{Maiolino_2008} combined a sample of $T_e$ derived low metallicity galaxies from \cite{Nagao_2006} with predictions from photoionization models in the high-metallicity regime. \cite{Bian_2018}\footnote{\cite{Bian_2018} did not include a strong-line calibration for R2.} constructed calibrations from a sample of local [OIII]$\lambda 4363$ emitters selected to match the location of $z \sim 2$ star-forming sources in the [NII]-BPT diagram \citep{Kewley_2013}. Finally, \cite{Nakajima_2022} extended the \cite{Curti_2017, Curti_2020} SDSS stacks to the extremely metal poor regime by including XMPGs identified from the EMPRESS survey \citep{Kojima_2020}. \cite{Nakajima_2022} further subdivided their calibrations characterized by high and low EW(H$\beta$) (i.e. EW(H$\beta$) $>200$~\AA\ and $<100$~\AA , respectively). Overall, the metallicity range for these calibrations differ, but we extrapolate each calibration over $6.9 \leq 12 + \log(\text{O/H} \leq 9.0)$. We indicate calibrated ranges as reported in the original papers as solid lines, whereas extrapolations as dotted lines in Figures \ref{fig:R2 Strong Line Comparison} - \ref{fig:R23 Strong Line Comparison}. We stress that extrapolating calibrations past their defined range can lead to nonphysical behaviours; however, we are extrapolating to examine the limitations of the calibrations. 

We determine the significance of deviation (in units of $\sigma$) for our JADES+ERO+CEERS sample to the predictions of each of the strong-line calibrations presented in Figures \ref{fig:R2 Strong Line Comparison} - \ref{fig:R23 Strong Line Comparison}. We determine the total deviation of our sample from the calibrations through a Monte Carlo technique. We evaluate the difference between our data points and the calibration values 10,000 times using values drawn randomly from normal distributions for the measured line ratios, metallicities, and calibrations. We include the line uncertainties, metallicity uncertainties, and the intrinsic dispersion of the calibrations ($\sigma_{cal}$) as the standard deviation for the respective distributions \footnote{\cite{Bian_2018} did not provide an estimate of the intrinsic dispersion for their calibrations. Following the procedure from \cite{Curti_2023}, we assume $\sigma_{cal} = 0.15$.}. We present in Table \ref{Table: Deviation} the total deviation between our sample and the respective calibrations. However, the sensitivity to metallicity varies over metallicity space for each strong-line diagnostic. For example, R23 has a weak dependence on metallicity at the turnaround point between $8.0 \lesssim 12+\log(\text{O/H}) \lesssim 8.5$, but a stronger dependence at lower metallicity ($12 +\log(\text{O/H}) \lesssim 7.65$). A primary concern for studies investigating the MZR is its slope, which is dependent upon how well the metallicities of galaxies, especially at the lower-mass end (lower metallicity), are determined. We therefore investigate how well each calibration does in predicting the $T_e$ derived oxygen abundances for each galaxy in our sample. We determine the offset between derived oxygen abundances by performing the same MC technique as above and then taking $12 +\log(\text{O/H})_{T_e}$ - $12 +\log(\text{O/H})_{cal}$. We present at the bottom of Figures \ref{fig:R2 Strong Line Comparison} - \ref{fig:R23 Strong Line Comparison} the offset to each respective calibration for our individual galaxies. The vertical lines represent the strong-line calibration failing for that object due to the calibration never reaching the measured line ratio at the given relation.  

\subsection{\textbf{R2}}

 There is approximately an order of magnitude scatter in the R2 ratio from Figure \ref{fig:R2 Strong Line Comparison}, suggesting there is notable diversity in the ISM conditions of our sample since R2 is highly dependent on the ionization parameter and hardness of ionizing spectrum. In comparison, we find a median R2 value of $-0.38$ with a standard deviation of $0.41$ while \cite{Cameron_2023}, using the same parent sample of this work but with selection criteria of $5.5 \leq z_{spec} \leq 9.5$ and S/N of H$\beta$ $\geq 5$, found a median R2 value of $-0.28$ with a standard deviation of $0.38$. We find the high-EW R2 calibration from \cite{Nakajima_2022} has the smallest significance of deviation to our sample with a $1.09\sigma$ deviation, though there are metallicity offsets over $\sim -0.5$ dex and 11 of our objects cannot be accounted for.
 
 R2 is rarely used in isolation, but is often employed to break degeneracies of other calibrations. However, for the high-\textit{z} Universe we clearly see there is significant scatter, thus suggesting the use of R2 as a degeneracy breaker in the high-\textit{z} Universe is problematic. We perform a Spearman correlation test on our JADES sample and find $\rho_{s} =0.58$ with a $p$-value of $0.001$, thus demonstrating a monotonic relationship with a low probability of an uncorrelated system reproducing the distribution. However, we see see similar R2 values across $\sim 1$~dex in metallicity. This insensitivity of R2 ratios to metallicity is possibly due to the ionization parameter-metallicity relation at these epochs, i.e., the ionization parameter-metallicity relation is not constant or has other dependencies \citep[e.g.,][]{Reddy_2023}. Overall, our sample demonstrates R2 is a poor metallicity diagnostic in the high-\textit{z} Universe, but the diversity in R2 values of our sample warrants a deeper investigation that is currently outside the scope of this paper.

\subsection{\textbf{O3O2}}

O3O2 also acts as a degeneracy breaker for other strong-line calibrations \citep{Maiolino_2019} as it primarily traces the ionization parameter with the metallicity dependence being secondary due to the ionization parameter-metallicity relation. We find a median O3O2 value of $1.08$ with a standard deviation of $0.36$, while \cite{Cameron_2023} found a median O3O2 value of $1.03$ with a standard deviation of $0.36$. Nearly our entire sample exhibits high O3O2 values with the smallest deviation calibrations ($0.96\sigma$) from \cite{Bian_2018} and the high-EW O3O2 calibration from \cite{Nakajima_2023} still failing to account for 22 of our galaxies and producing metallicity offsets $\sim 0.6$ dex. 
 
We find a Spearman correlation of $\rho_{s} =-0.44$ with a $p$-value of $0.02$, thus demonstrating a correlation, albeit weak. However, we find similar O3O2 values across $\sim 1$~ dex in metallicity similar to R2. Therefore, although our sample is small, this finding suggests that O3O2 is neither a good O/H diagnostic nor an appropriate degeneracy breaker for other strong-line diagnostics in the high-\textit{z} Universe. A more detailed picture of O3O2 was presented by \cite{Cameron_2023}, in which they compared O3O2 against R23 (their Figure 5), which is ultimately comparing tracers of ionization parameter and total excitation, respectively. \cite{Cameron_2023} found the JADES sample to exhibit much higher O3O2 values at a given R23 value compared to $z \sim 2$ MOSDEF galaxies, which already traced the extremes of SDSS $z \sim 0 $ populations. \cite{Cameron_2023} concluded that galaxies across the sample exhibit very high ionization parameters. This high ionization is reflected in Figures \ref{fig:O3O2 Strong Line Comparison} and Table \ref{Table: Deviation} as the majority of calibrations fail to return a $12 + \log(\text{O/H})$ value at O3O2 ratios we measure. An explanation for this high ionization would be simple if our sample had lower O/H values since that would suggest the ionization-metallicity relation is constant. However, ionization is generally higher at fixed metallicity in our sample, thus suggesting a physically-driven change, though a full characterization will be explored in forthcoming work.
 
\subsection{\textbf{R3}} \label{R3}

In contrast to R2 and O3O2, we see little scatter in our sample for R3. We find a median R3 value of $0.81$ with a standard deviation of $0.12$. \cite{Cameron_2023} also found a median R3 value of $0.74$ with a standard deviation of $0.86$. We find the calibration from \cite{Bian_2018} has the smallest significance of deviation for our sample with a $0.50\sigma$ deviation, though four of our galaxies cannot be predicted by the calibration, metallicity offsets are up to $\sim -0.6$ dex, and we are ultimately comparing against the extrapolation. Nonetheless, the R3 calibration from \cite{Bian_2018} best traces our sample out of the local calibrations.

We find a Spearman correlation of $\rho_{s} =0.62$ with a $p$-value of $0.0004$, thus demonstrating there is still a strong relationship between R3 and metallicity. However, R3 has a characteristic turnover locally, which requires identifying which of the two branches applies. Interestingly, we see an apparent flattening of our sample across the double-valued R3 sequence. R3 is similar to R2 in that it is highly degenerate with the ionization parameter, the hardness of the ionizing spectrum, and the relation between metallicity and ionization parameter \citep{Kewley_2008, Maiolino_2019}. As such, the flattening of our objects across the double-valued sequence, in addition to the large scatter in R2 and O3O2, suggests significant ionization across $\sim 1~\text{dex}$ in metallicity in our sample. Without probing higher metallicities it is difficult to conclude whether the characteristic turnover is present in the high-\textit{z} Universe. If R3 is confirmed to have a minimal turnover then R3 as a metallicity diagnostic is not viable in the high-\textit{z} Universe. Overall, forthcoming work will investigate whether R3 turns over and the origins of the excess R3 values.  

\subsection{\textbf{R23}}

R23 is the most widely used strong-line calibration in determining metallicity because, unlike R2 and R3, R23 is an indication of the total excitation of a galaxy as it combines the different ionization states of oxygen. There is still a high dependence on the ionization parameter, however, along with a double branching that requires employing other strong-line diagnostics, such as R2 or O3O2, to break the degeneracy. R23 has already been employed in the high-\textit{z} Universe \citep[e.g.,][]{Nakajima_2023}; however, we find moderate deviation from our sample for the R23 calibrations. Specifically, we find the calibration from \cite{Bian_2018} to have the smallest significance of deviation to our sample with a $0.58\sigma$ deviation, though as can be seen in Figure \ref{fig:R23 Strong Line Comparison}, the majority of our points do not fall within the calibrated range of \cite{Bian_2018}, 24 of our galaxies cannot be predicted, and metallicity offsets up $\sim 0.5$~dex exist. From Figure \ref{fig:R23 Strong Line Comparison}, however, we see visually the large EW sample from \cite{Nakajima_2022} best traces the upper envelope of our objects for a calibrated range, though metallicity offsets range between $\sim -0.5$ and $0.5$ dex.  We find a median R23 value of $0.97$ and a standard deviation of $0.13$. \cite{Cameron_2023} also found a median R23 value of $0.90$ with a standard deviation of $0.10$. Overall, the R23 ratios of our JADES sample suggests significant excitation across $\sim 1 \text{dex}$ in metallicity than what is typically seen in local galaxies.

It is clear that a self-consistent calibration of R23 is needed for the high-\textit{z} Universe, but it is difficult to conclude whether R23 is appropriate for the high-\textit{z} Universe. We find a Spearman correlation of $\rho_{s} =0.68$ with a $p$-value of $4.4 \times 10^{-5}$, which indicates there is a strong correlation of R23 with metallicity. However, similar to our R3 ratios, we cannot determine whether R23 turns over or not. We cannot probe past the low-\textit{z} turnover point ($8.0 \lesssim 12+\log(\text{O/H}) \lesssim 8.5$) with our limited sample, but visually and with the Spearmen Rank correlation/p-value, the metallicity dependency of R23 is possibly inadequate for a high-\textit{z} metallicity indicator, especially if this trend continues past the low-\textit{z} turnover point. A stacking procedure, similar to \cite{Curti_2017, Curti_2020}, is necessary to probe past the low-\textit{z} turnover point. 

\subsection{\textbf{Comparison to High-\textit{z} Calibration}}

In addition to a high-\textit{z} [OIII]$\lambda 4363$ sample, \cite{Sanders_2023} provided the first high-\textit{z} strong-line calibrations. Accordingly, we compare their calibrations for R2, O3O2, R3, and R23 to our sample in Figures \ref{fig:R2 Strong Line Comparison} - \ref{fig:R23 Strong Line Comparison}. We determine the significance of deviation as described in Section \ref{Comparison to Locally Derived Strong Line Calibrations} for each calibration from \cite{Sanders_2023}. We find our sample to be  $1.24\sigma$, $1.17\sigma$, $0.77\sigma$ $0.81\sigma$ away for R2, O3O2, R3, and R23, respectively. The R3 and R23 calibrations from \cite{Sanders_2023} do visually trace the upper envelope of our sample where other local calibrations underestimate. However, at $12 + \log(\text{O/H}) \lesssim 8.0$ the extrapolation of the calibration from \cite{Bian_2018} predicts higher R3 ratios at a given metallicity than \cite{Sanders_2023}, thus leading to the higher deviation reported for the \cite{Sanders_2023} calibration. For R2 and O3O2, the deviations reported for the \cite{Sanders_2023} calibration are due to the significant scatter in our sample. 

We note here and demonstrate in the Appendix A that there would be a systematic offset introduced when comparing a calibration and a sample with different metallicity prescriptions(e.g., \texttt{Pyneb} and \cite{Izotov_2006}, thus emphasizing the importance of self-consistency before systematics $T_e$ choice are better constrained. Nonetheless, the high-\textit{z} calibration from \cite{Sanders_2023} visually traces our sample well in the strong-lines investigated in the current work, but as discussed in Section \ref{Discussion}, larger [OIII]$\lambda 4363$ samples are clearly needed for future high-\textit{z} Universe strong-line calibrations.

\subsection{\textbf{Photoionization Models}} \label{Photoionization Models}

A common alternative to determining metallicities through the  T$_e$ method or strong-line calibrations is the use of photoionization models due to the range of properties that can be explored \citep[e.g.,][]{Tremonti_2004, Perez_2014, Dopita_2016, Vale_2016}. However, this approach is currently limited as it is difficult to capture the complexity of HII regions and a number of assumptions are employed (e.g., plane-parallel atmospheres, the ionizing spectrum, and dust depletion) \citep{Maiolino_2019}. This area has improved with certain frameworks introducing Bayesian approaches where multiple emission lines are used to identify the best corresponding model returned from a grid (e.g., \texttt{PyNeb} \citep{Luridiana_2015}, \texttt{CLOUDY} \citep{Ferland_2013}, etc.) while minimizing assumptions. One such code is \texttt{HII-CHI-Mistry} from \cite{Perez_2014}. 

\cite{Perez_2014} used the synthesis spectral code \texttt{CLOUDY} v13.03 \citep{Ferland_2013} using POPSTAR \citep{Molla_2009} stellar evolutionary models assuming an instantaneous burst with an age of 1 Myr with an initial mass function from \cite{Chabrier_2003}. They range the ionization parameter between $-1.50 \leq \log(\text{U}) \leq -4.00$ in steps of 0.25 dex, the oxygen abundance between $7.1 \leq 12+\log(\text{O/H}) \leq 9.1$ in steps of 0.1 dex, and consider variations in the N/O ratio between $0.0 \leq \text{N/O} \leq -2.0$ in steps of 0.125 dex, thus totaling $3927$ models. It would be excessive to compare all the models, so we compare against the full metallicity range for N/O values of -2.0 (purple), -1.0 (green), and 0.0 (red) and $\log(\text{U})$ values of -1.5 (dashed), -2.5 (solid), and -3.5 (dotted).

\begin{figure}[hbt!]
    \centering
    \includegraphics[width = \columnwidth]{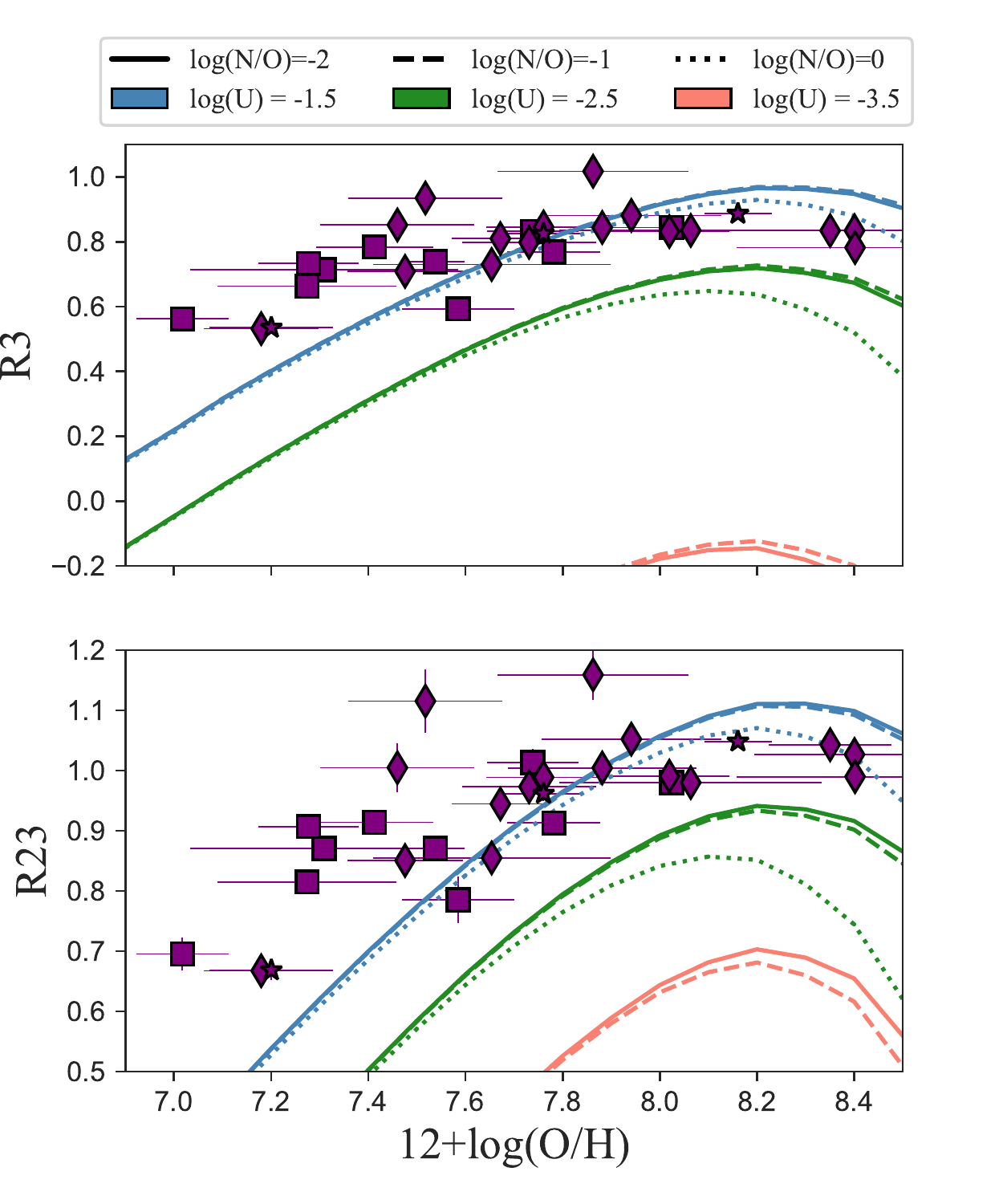}
    \caption{R3 vs $12 + \log(\text{O/H})$ and R23 vs $12 + \log(\text{O/H})$ with photoionization models from \cite{Perez_2014}. Various line styles and colors represent different nitrogen/oxygen abundance ratios and ionization parameters, respectively. Symbols are the same as in Figure \ref{fig:R2 Strong Line Comparison}. }
    \label{fig:Perez_Montero_comparison}
\end{figure}

We present in Figure \ref{fig:Perez_Montero_comparison} our JADES sample and the grid models returned from \cite{Perez_2014}. Our JADES sample is best traced by the log(U) $= -1.5$ models, though our most metal-poor galaxies require a higher ionization parameter while our least metal-poor galaxies fall close to log(U) = $-2.5$ models. The N/O models are indistinguishable as the values converge for our sample range. As such, it is still unclear whether we are dealing with extremely nitrogen poor systems. Nitrogen enrichment could be moderate yet exist in higher ionization states that we are unable to probe with [NII]. As mentioned, \cite{Cameron_2023} found no detections of nitrogen even with 7 hour deep G395M/F290LP spectra, indicating future difficulty in examining N/O abundance ratios in metal-poor galaxies. Yet, GN-z11 revealed rarely-seen NIV]$\lambda 1486$ and NIII]$\lambda 1748$ lines \citep{Bunker_2023}, with subsequent explanations implying unusually high N/O abundance \citep{Cameron_2023_gnz11, Senchyna_2023}. N/O trends at high-\textit{z} are outside the scope of the current work, but our JADES sample demonstrates the importance constraining N/O trends in the high-\textit{z} Universe and how nitrogen is handled in photoionization models.

\subsection{\textbf{A new projection in the R2-R3-O/H space}} \label{Mirko's New Calibration}

The set of calibrations presented by \cite{Sanders_2023} (in particular those related to the R3 and R23 diagnostics) are starting to provide a more accurate representation of the distribution of galaxies with direct metallicities in the high-z Universe. 
Nonetheless, the calibration curves are still poorly sampled at both the low- and high-metallicity end, with the majority of galaxies with $T_e$ measurements distributed within the $7.6<12+\log(\text{O/H})<8.2$ abundance range, close to the plateau of the calibrations.
Moreover, given the relatively high-excitation properties of these sources (which boosts R3 and R23 at fixed O/H), the slope of the calibration curves appears to flatten further compared to most of the low-z calibrations, the plateau is hence wider, and the dynamic range in which these line ratios are sensitive to a variation in metallicity is reduced: this means that, for instance, at a value of R3 $=0.8$ (above which more than $50$ per cent of the currently available calibration sample resides) the `gap' between the low- and high-metallicity solutions of the calibration is $\sim0.6$~dex. 

Here, we attempt to provide a novel calibration based on a similar sample as described in \cite{Nakajima_2022}, but that however involves a different projection in the space defined by log([OII]$\lambda 3727,29$/H$\beta$), log([OIII]$\lambda 5007$/H$\beta$), and metallicity.
More specifically, such new diagnostic, which we here label as \^R, is defined as 
$\text{\^R} = 0.47\ \text{R2} + 0.88\ \text{R3}$. As described more in detail in Appendix B, such linear combination corresponds to a rotation of $61.82$ degrees around the O/H-axis in the R2-R3-O/H space, a projection that minimizes the scatter of our calibration sample in \^R at fixed metallicity over the full O/H range spanned by the galaxy calibration sample.
We fit a fourth order polynomial to the \^R vs O/H relation as shown in Figure~\ref{fig:new_calibration}, with the best-fit coefficients that are provided in Appendix B.
Compared to R23, this diagnostic has a wider dynamic range in its low-metallicity branch, spanning an interval of values between $-0.2$ and $0.8$ between $7.0<$12+log(O/H)$<8.0$, and shows a narrower turnover and plateau region.

We compare our observed JWST sample with the \^R diagnostic in Figure~\ref{fig:new_calibration}. We find a reasonably good agreement between \^R-predicted and observed metallicities for the high-z sample, with no systematic offset above or below the calibration curve: the points scatter around the best-fit relation with a median offset in \^R of 0.002~dex at fixed O/H, a median absolute deviation of 0.13~dex, a dispersion of 0.19~dex, and a significance of $1.00\sigma$\footnote{The dispersion of the \^R calibration is lower than all local calibrations. A lower intrinsic dispersion can increase the significance of deviation since the calibration varies less compared to a calibration with higher dispersion that is able to ``roam" closer to more distant points when performing a Monte Carlo procedure.} given an intrinsic dispersion of the calibration of $0.058$~dex.

\begin{figure}[hbt!]
    \centering
    \includegraphics[width = \columnwidth]{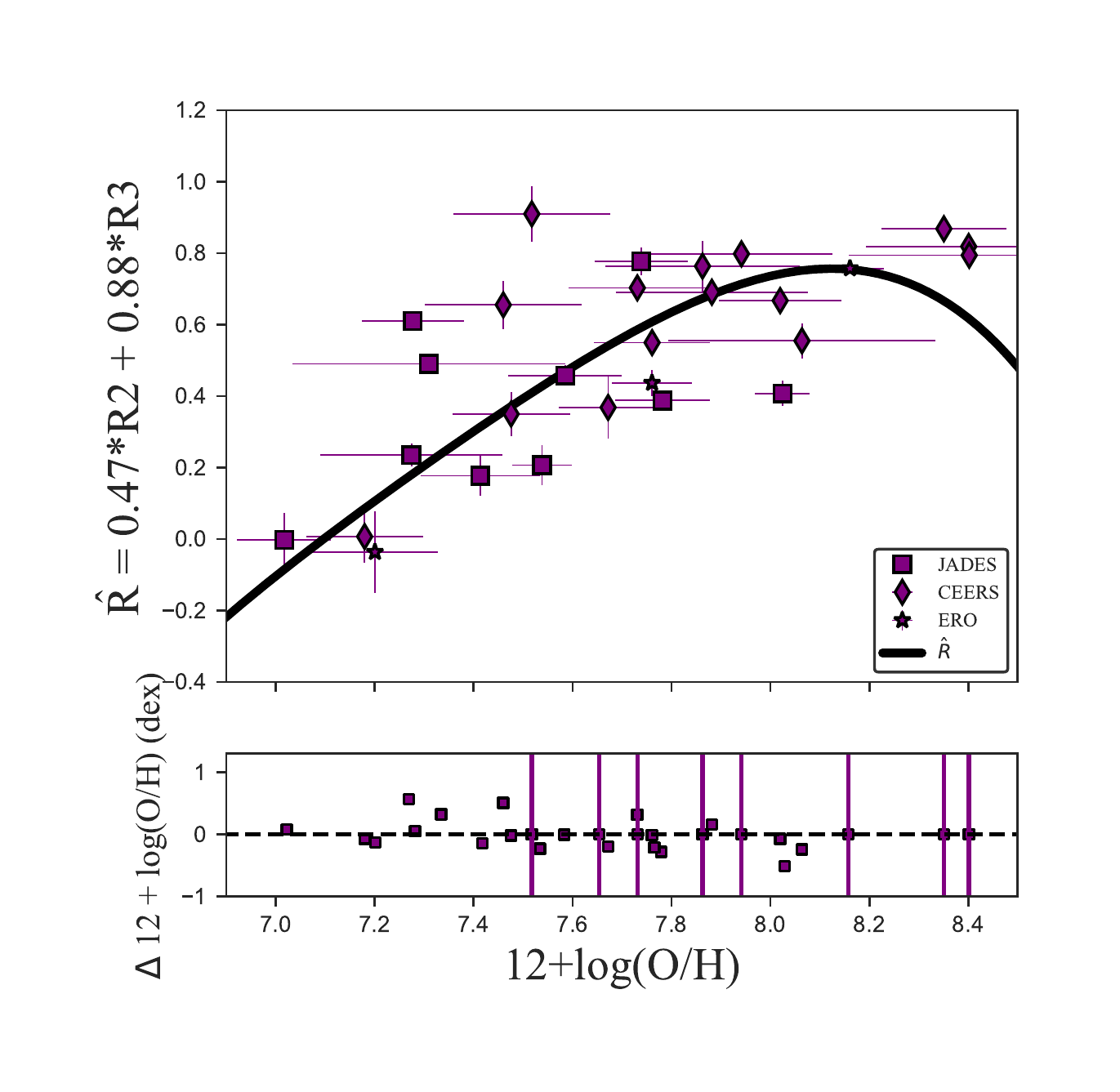}
    \caption{The JWST sample with auroral lines measurements analysed in this work is compared against the \^R diagnostic presented in Section~\ref{Mirko's New Calibration}.
    Symbols are the same as in Figure~\ref{fig:R2 Strong Line Comparison}. The high-z sample with $T_e$ metallicities is predicted by the \^R calibration with a median absolute offset of 0.13~dex and a standard deviation of 0.19~dex.}
    \label{fig:new_calibration}
\end{figure}

\section{\textbf{Discussion}} \label{Discussion}

\subsection{\textbf{Strong-line Diagnostics}} \label{Strong-line Diagnostics}

From Figures \ref{fig:R2 Strong Line Comparison} - \ref{fig:R23 Strong Line Comparison} and Table \ref{Table: Deviation}, we see clear discrepancies between locally-derived strong-line calibrations and our JADES sample. We find that a single calibration cannot simultaneously account for all galaxies across all diagnostics. 

The largest discrepancies between local-calibrations and our JADES sample are for the R2 and O3O2 diagnostics, which is most likely caused by R2 and O3O2 being insensitive to metallicity at these redshifts, i.e., R2 and O3O2 are not appropriate metallicity indicators or degeneracy breakers for the high-\textit{z} Universe. Recently, \cite{Reddy_2023} concluded that electron gas density potentially has a larger responsibility than metallicity in modulating the ionization parameter in these early epochs. We are potentially observing this result in Figures \ref{fig:R2 Strong Line Comparison} and \ref{fig:O3O2 Strong Line Comparison} where we have consistently high ionization ratios over our metallicity space, but further investigation is needed.

For our sample, R3 and R23 still indicate a dependency on metallicity at these high redshifts. Spearman correlations of $\rho_{s}$ = $0.62$ \& $0.68$ with $p$-values of $0.0004$ \& $4.4 \times 10^{-5}$, respectively, further corroborate this finding. However, we do observe flattening of our sample compared with local R3 and R23 calibrations, possibly suggesting future difficulty when applying these diagnostics in the high-\textit{z} Universe, especially at moderate metallicites. This flattening is potentially a result of an evolution in the ionization parameter-metallicity relation that has a higher dependency on electron densities \citep{Reddy_2023}, though a much more detailed analysis on a larger sample size of high-\textit{z} [OIII]$\lambda 4363$ emitters and stacked spectra of several hundreds of galaxies to probe to higher metallicities ($12 + \log(\text{O/H}) \gtrsim 8.0 - 8.5$) is required to examine the physical origins and establish whether there is a turnover for R3 and R23.     

Overall, any local calibration for R2, O3O2, R3, and R23 clearly fails to simultaneously match our sample: There is a clear need for a self-consistent revision of the calibrations in the high-\textit{z} Universe using JWST, and we caution against the use of locally derived calibrations being applied to high-\textit{z} Universe. We postpone deriving new R2, O3O2, R3, and R23 calibrations for the high-\textit{z} Universe as our sample is limited and it is best to remain self-consistent until systematics between spectroscopic reduction pipelines are better characterized. As such, it is essential to continue constructing samples of [OIII]$\lambda 4363$ in the high-\textit{z} Universe with JWST. 

While [OIII]$\lambda 4363$ sample sizes increase and calibrations improve, the \^R projection presented in this paper and the high-\textit{z} calibrations from \cite{Sanders_2023} provide the best match to high-\textit{z} [OIII]$\lambda 4363$ derived metallicities.

\subsection{\textbf{EW$_0$(H$\beta$) Discrepancies}}

Rest-frame EWs(H$\beta$) can range between $10 - 600$~\AA~for [OIII]$\lambda4363$ emitters \citep[e.g.,][]{Maiolino_2019, Izotov_2021b, Laseter_2022, Nakajima_2022}. As such, when \cite{Nakajima_2022} were developing their calibrations they investigated whether the accuracy of strong-line diagnostics could be improved if one includes rest-frame EWs(H$\beta$) as an additional parameter. This investigation lead \cite{Nakajima_2022} to separate calibrations over the rest-frame EWs(H$\beta$) range of $20 \text{~\AA} \lesssim \text{EW}_0(\text{H}\beta) \lesssim 300 \text{~\AA}$ as we have shown in Figures \ref{fig:R2 Strong Line Comparison} - \ref{fig:R23 Strong Line Comparison}. Therefore, their high EW fit (EW$_0$(H$\beta$) $\geq 200$~\AA) is based on the most extreme EW$_0$(H$\beta$) objects in their calibration sample. It is thus warranted to determine the rest-frame EWs of H$\beta$ for our JADES galaxies and examine their strength. 
  
To determine EW$_0$(H$\beta$) for our JADES objects we interpolate the best-fit continuum to our PRISM data from \texttt{PPXF} over a $60$~\AA~bin around the H$\beta$ line center in our R1000 data, divide the measured flux of H$\beta$ from the R1000 fits by the interpolated best-fit continuum, and then divide by $(1+z)$. We include EW$_0$(H$\beta$)~\AA~ for our objects in Table \ref{Table: Galaxy Properties}. 

Although our JADES sample demonstrates excitation ratios higher than any local R3 and R23 calibration (excluding the extrapolation of \cite{Bian_2018}), the high EW$_0$ calibration (EW$_0$(H$\beta$) $>200$~\AA) from \cite{Nakajima_2022} lies closest to the upper envelope of our sample. However, we find the median EW$_0$(H$\beta$) for our JADES sample to be $\sim 170$~\AA, with the minimum being $\sim 70$~\AA~ and the max being $\sim 550$~\AA. Interestingly, we find the median EW$_0$(H$\beta$) becomes $\sim 120$~\AA~when excluding galaxies in our sample beneath $z = 4.0$. As such, there is an apparent decrease in rest-frame EWs(H$\beta$) of high-\textit{z} [OIII]$\lambda 4363$ emitters compared to local metal-poor objects with [OIII]$\lambda 4363$ detections, even though we find higher ionization/excitation ratios for our sample.

An increase in the luminosity of [OIII]$\lambda 4363$ in the high-\textit{z} Universe could account for the EW$_0$(H$\beta$) disparity in that galaxies in earlier epochs have intrinsically brighter [OIII]$\lambda 4363$ at a fixed EW$_0$(H$\beta$). However, it is difficult to characterize whether there is a physically driven increase in the luminosity of [OIII]$\lambda 4363$ for our sample due to limited $z > 1$ [OIII]$\lambda 4363$ samples, lack of flux calibrations for most studies, and undetermined mass completion limits. Nonetheless, a line luminosity increase is expected due to the FMR. At lower metallicities and/or masses we expect an increase in the SFR, and thus luminosity. However, it is debated whether the FMR evolves with redshift, though \cite{Curti_2023_MZR}, using the same parent data set as the current work, demonstrates galaxies sit preferentially below local FMR predictions with increasing redshift ($z \gtrsim 6$), such that these galaxies are significantly less enriched at a given SFR and stellar mass. 

In general, [OIII]$\lambda 4363$ would be more luminous with an increase in sSFR and/or a decrease in metallicity. However, we would expect an increase in sSFR to be associated with higher rest-frame EWs(H$\beta$) relative to local counterparts, but for our JADES objects we find rest-frame EWs(H$\beta$) lower than local galaxies that have reduced ionization/excitation ratios at similar metallicities compared to our sample. Therefore, we expect the [OIII]$\lambda 4363$ luminosity of our JADES sample to be driven by lower metallicities, thus reflecting a number of possible processes such as pristine gas accretion \citep{Mannucci_2010} and efficient metal removal from stellar winds  that are expected to increase with a top-heavy IMF \citep{Palla_2020}. However, as mentioned, \cite{Cameron_2023} found our parent sample exhibits excitation ratios resembling extreme star-formation galaxies, such as blueberries \citep{Yang_2017a} and blue compact dwarf galaxies \citep{Sargent_1970, Cairos_2010C} that are known to have high sSFRs ($10^{-7} \text{yr$^{-1}$} \lesssim \text{sSFR} \lesssim 10^{-8} \text{yr$^{-1}$}$). In addition, \cite{Curti_2023_MZR} found our parent sample occupies the same region of the MZR as these extreme star-forming galaxies.

Overall, the picture is opaque. It is peculiar that we are simultaneously observing galaxies with lower rest-frame EWs(H$\beta$) and higher excitation values relative to local analogs that have high sSFRs. In addition, a number of possible processes, such as an evolving FMR, variations in metal-cooling due to elemental production time scales (e.g., oxygen being enriched rapidly due to the production from core-collapse supernovae, compared to similar cooling curves from nitrogen and carbon that are enriched by massive stars and type Ia supernovae), or more extreme, poorly understood thermal and density structure variations in the emitting nebulae \citep{Cameron_2022, Reddy_2023}, could all affect the luminosity of [OIII]$\lambda 4363$, metallicity determinations, and ionization/excitation values. In addition, \cite{Reddy_2023} proposed that electron density plays a larger role in regulating the ionization parameter, which in return would affect the temperature distribution of HII regions where [OIII]$\lambda 4363$ originates from. Our sample clearly demonstrates the necessity for a deeper investigation into the production of [OIII]$\lambda 4363$ in the high-\textit{z} Universe.

\section{\textbf{Summary and Conclusions}} \label{Conclusions}

We have identified 10 [OIII]$\lambda 4363$ detections discovered from ultra-deep JWST/NIRSpec MSA spectroscopy from the JADES DEEP survey, which is only a small fraction of the final JADES spectroscopic dataset . We applied the $T_e$-method to determine gas-phase oxygen abundances to examine how well local strong-line calibrations match a robust high-\textit{z} [OIII]$\lambda 4363$ sample. Our main findings are summarised as follows:

\begin{enumerate}
  \item The local strong-line metallicity calibrations investigated do not provide good simultaneous predictions for the metallicities across our sample as seen in Figures \ref{fig:R2 Strong Line Comparison} - \ref{fig:R23 Strong Line Comparison}. Specific calibrations have smaller deviations for various diagnostics while completely failing for lower metallicity galaxies, thus demonstrating the necessity for a systematic re-calibration of R2, O3O2, R3, and R23 strong-line diagnostics in the high-\textit{z} Universe. We caution against employing locally derived calibrations in the high-z Universe.

  \item There is weak correlation between R2 and O3O2 with metallicity. If larger samples with higher metallicity galaxies support this finding then R2 and O3O2 would be inadequate diagnostics for deriving metallicities or breaking degeneracies in the high-\textit{z} Universe. There is also an order of magnitude scatter at fixed metallicity in our sample for R2 and O3O2 diagnostics, further demonstrating ISM diversity that is potentially diminishing the dependency of R2 and O3O2 with metallicity. R3 and R23 correlate with metallicity, but elevated, comparable line-ratios across $\sim 1$~dex in metallicity demonstrates a flattening of the strong-lines with metallicity. If this trend continues past the turnover point between $8.0 \lesssim 12 + \log(\text{O/H}) \lesssim 8.5$ then R3 and R23 would be problematic to use in the high-\textit{z} Universe as metallicity would be indistinguishable without a substantial degeneracy breaker.

  \item The new \^R projection (\^R = 0.47 R2 + 0.88 R3) and high-\textit{z} calibrations (R3 \& R23) from \cite{Sanders_2023} provide the best match to our sample overall. However, larger high-\textit{z} [OIII]$\lambda 4363$ sample sizes are needed that extend to higher metallicities past the plateaus of the calibrations. 

  \item The rest-frame H$\beta$ EWs of our JADES sample are moderate with the median being $\sim 170$~\AA. However, excluding galaxies lower than $z = 4$ in our JADES sample yields a median of $\sim 120$~\AA, which contrasts local galaxies with rest-frame EWs(H$\beta$) $\sim 300$~\AA~used to derive local calibrations that still fall beneath the ionization/excitation ratios of our sample. In addition, our elevated excitation values, along with the findings of \cite{Cameron_2023} and \cite{Curti_2023_MZR}, demonstrates our sample closely matches extreme star-formation galaxies, such as blueberries \citep{Yang_2017a} and blue compact dwarf galaxies \citep{Sargent_1970, Cairos_2010C} that are known to have some of the highest sSFRs ($10^{-7} \text{yr$^{-1}$} \lesssim \text{sSFR} \lesssim 10^{-8} \text{yr$^{-1}$}$). The combination of these findings does not present a clear description of [OIII]$\lambda 4363$ production in the high-\textit{z} Universe, thus warranting a much deeper examination into the possible processes.
  
  %, though larger, self-consistent samples of [OIII]$\lambda 4363$ are needed. 

\end{enumerate}

\section{Acknowledgments}
This material is based upon work supported by the National Science Foundation Graduate Research Fellowship under Grant No. 2137424. ECL acknowledges support of an STFC Webb Fellowship (ST/W001438/1). S.C acknowledges support by European Union’s HE ERC Starting Grant No. 101040227 - WINGS. AJC acknowledges funding from the "FirstGalaxies" Advanced Grant from the European Research Council (ERC) under the European Union’s Horizon 2020 research and innovation programme (Grant agreement No. 789056). R.M. and W.B. acknowledge support by the Science and Technology Facilities Council (STFC) and by the ERC through Advanced Grant 695671 "QUENCH". RM also acknowledges funding from a research professorship from the Royal Society. AJB, AJC, JC, IEBW, AS and GCJ acknowledge funding from the "FirstGalaxies" Advanced Grant from the European Research Council (ERC) under the European Union's Horizon 2020 research and innovation programme (Grant agreement No. 789056). S.A. and B.R.P acknowledge support from the research project PID2021-127718NB-I00 of the Spanish Ministry of Science and Innovation/State Agency of Research (MICIN/AEI). JWST/NIRCam contract to the University of Arizona NAS5-02015. DJE is supported as a Simons Investigator and by JWST/NIRCam contract to the University of Arizona, NAS5-02015. Funding for this research was provided by the Johns Hopkins University, Institute for Data Intensive Engineering and Science (IDIES). RS acknowledges support from a STFC Ernest Rutherford Fellowship (ST/S004831/1). BER acknowledges support from the NIRCam Science Team contract to the University of Arizona, NAS5-02015. The authors acknowledge use of the lux supercomputer at UC Santa Cruz, funded by NSF MRI grant AST 1828315. The research of CCW is supported by NOIRLab, which is managed by the Association of Universities for Research in Astronomy (AURA) under a cooperative agreement with the National Science Foundation. This research is supported in part by the Australian Research Council Centre of Excellence for All Sky Astrophysics in 3 Dimensions (ASTRO 3D), through project number CE170100013. B.R.P. acknowledges support from the research project PID2021-127718NB-I00 of the Spanish Ministry of Science and Innovation/State Agency of Research (MICIN/AEI). J.S. acknowledges support by the Science and Technology Facilities Council (STFC), ERC Advanced Grant 695671 "QUENCH".

\newpage
\clearpage
\bibliographystyle{aasjournal}
\bibliography{bibliography.bib}

%\clearpage

\appendix

\section{Metallicity Prescriptions and Collisional Strengths}
\label{appendix:Pyneb vs Izotov}

\begin{figure}[hbt!]
    \centering
    \textbf{Izotov et al. (2006) Derived Abundances}\par
    \includegraphics[width = 0.75\columnwidth]{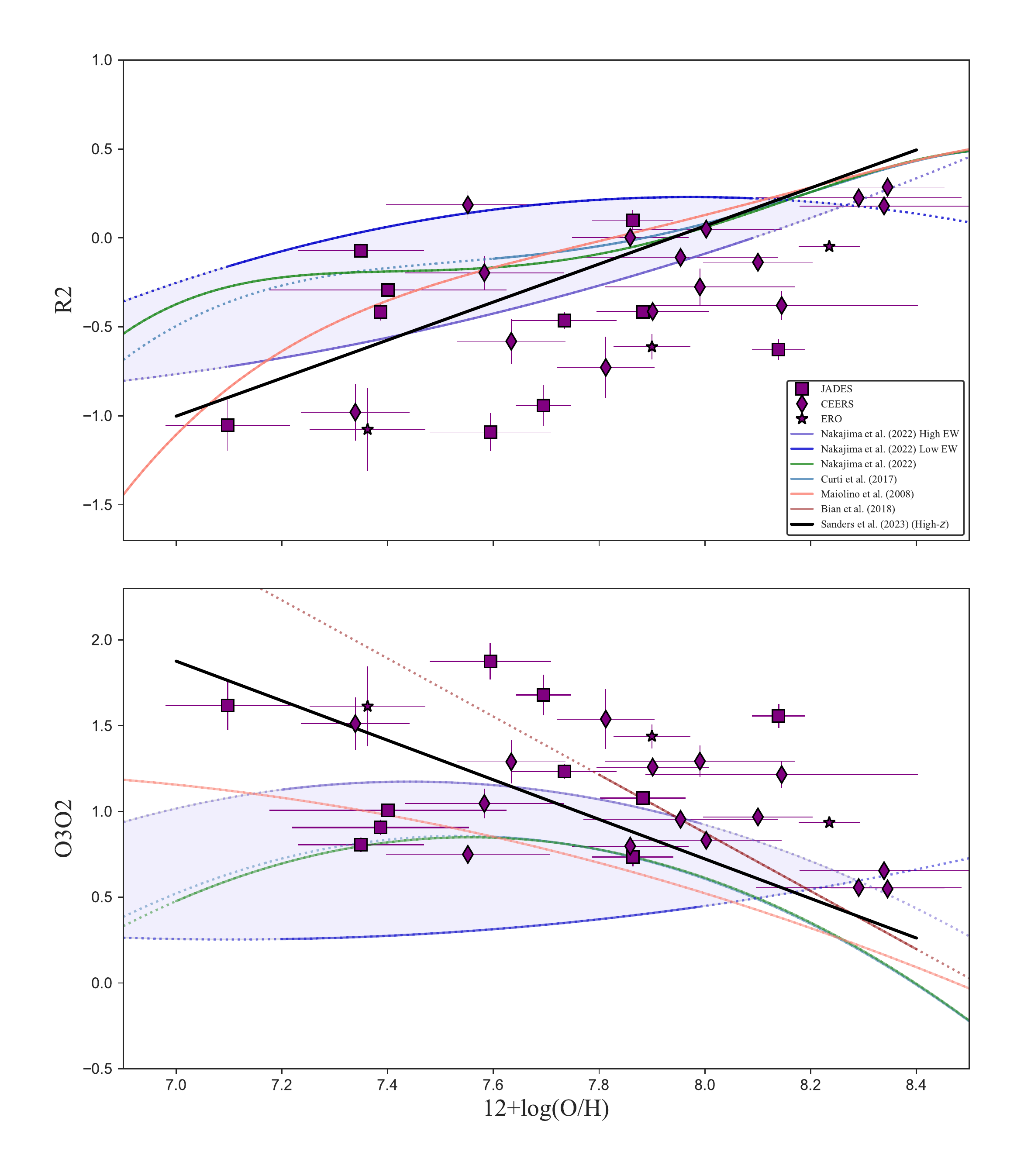}
    \caption{Our sample with derived abundances using \cite{Izotov_2006} compared against local calibrations for R2 and O3O2, respectively. }
    \label{fig:R2_O3O2_Izotov_comp}
\end{figure}

\begin{figure}[hbt!]
    \centering
    \includegraphics[width = 0.75\columnwidth]{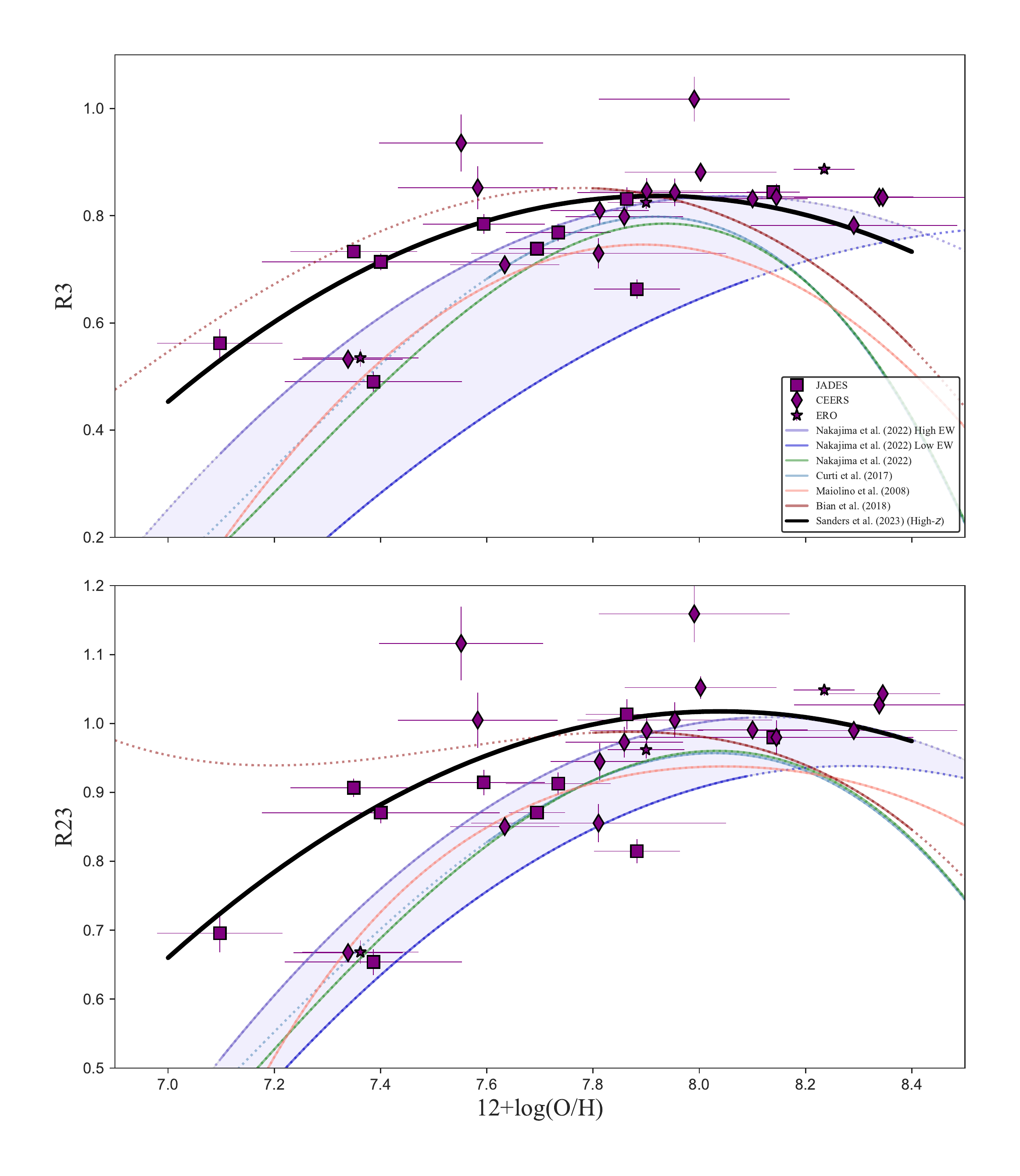}
    \caption{Same as Figure \ref{fig:R2_O3O2_Izotov_comp} except for R3 and R23. }
    \label{fig:R3_R23_Izotov_comp}
\end{figure}

In Section \ref{Metallicity Prescription Choice} we examined systematic offsets introduced when changing metallicity prescriptions between \texttt{PyNeb} and the empirical relations from \cite{Izotov_2006}. We demonstrated there is a $-0.11$~dex offset between \texttt{PyNeb} and \cite{Izotov_2006} for our sample. The median error of \texttt{PyNeb} derived abundances for our sample is $0.12$~dex, so the systematics introduced when choosing a metallicity prescription are comparable to the associated error with our measurements. We demonstrate these systematics further in Figures \ref{fig:R2_O3O2_Izotov_comp} and \ref{fig:R3_R23_Izotov_comp} by deriving metallicities for our sample using the \cite{Izotov_2006} prescription and comparing against strong-line calibrations as we did in Figures \ref{fig:R2 Strong Line Comparison} - \ref{fig:R23 Strong Line Comparison}. We clearly see our sample more closely matches R3 and R23 local calibrations when using \cite{Izotov_2006}. However, more recent local calibrations, such as \cite{Curti_2017, Curti_2020} and \cite{Nakajima_2023}, along with the high-\textit{z} calibrations from \cite{Sanders_2023}, employed \texttt{PyNeb} for their $T_e$ determinations and therefore their calibrations. As such, if we were to determine the respective \texttt{PyNeb} calibrations using \cite{Izotov_2006} instead then the main findings of the paper remain. It is clear that choosing a metallicity prescription matters, and thus future studies combining multiple samples should consistently re-derive metallicities for each respective sample to remain self-consistent.

\begin{figure}[hbt!]
    \centering
    \includegraphics[width = 0.75\columnwidth]{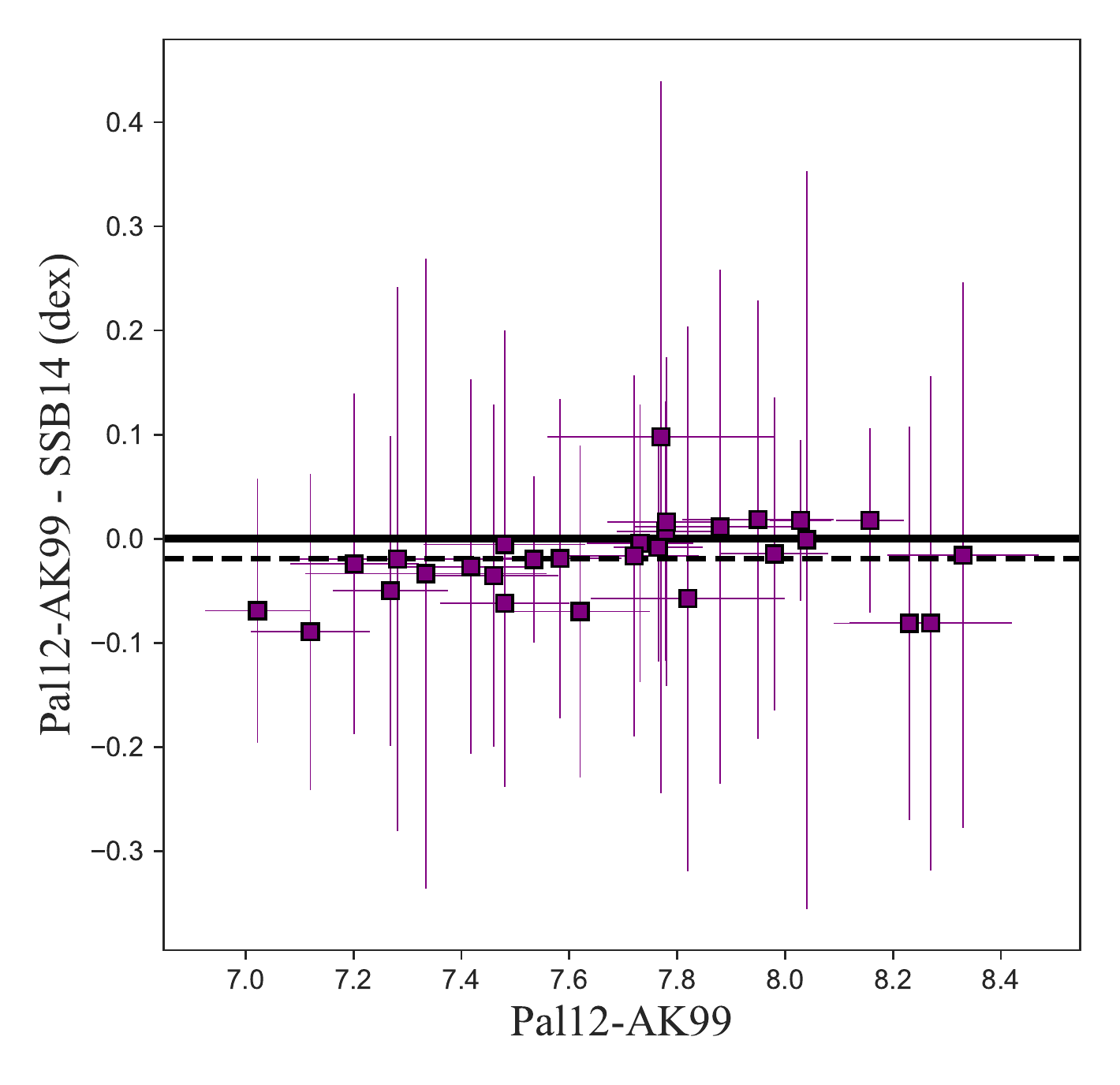}
    \caption{Deviation between metallicites derived by \texttt{PyNeb} using O2+ collision strengths from \cite{AK_1999} \& \cite{Palay_2012} and \cite{Storey_2014}. The solid line represents unity, whereas the dashed line represents the median offset between \cite{AK_1999} \& \cite{Palay_2012} and \cite{Storey_2014}.}
    \label{fig:collision_comp}
\end{figure}

In addition to metallicity prescription choice, the atomic data used, such as the options provided in \texttt{PyNeb} or the \texttt{CLOUDY} configurations used in \cite{Izotov_2006}, can introduce systematic offsets. For example, we use the O2+ collision strengths from \cite{AK_1999} \& \cite{Palay_2012} when determining our metallicities, but \cite{Sanders_2023} used O2+ collision strengths from \cite{Storey_2014} (the default of \texttt{PyNeb}) when deriving their metallicities, hence why we re-derived metallicities from \cite{Sanders_2023} for our sample. 
%Consequently, we are therefore comparing a calibration to metallicities derived with different O2+ collisional strengths.
Similar to Figure \ref{fig:Systematics}, we present in Figure \ref{fig:collision_comp} the systematic offsets in metallicity for our sample introduced when choosing to use O2+ collision strengths between \cite{Storey_2014} and \cite{AK_1999} \& \cite{Palay_2012} internal to \texttt{PyNeb}. We find a median metallicity offset of $0.02$~dex, but there are offsets between $\sim -0.1$ and $0.1$~dex in our sample. It is clear the systematic offsets between metallicity prescriptions are overall larger, but the offsets introduced when choosing collisional strengths can be non-negligible. 

Overall, it is clear that choosing a metallicity prescription, and to a lesser extent the collisional strengths, matters. The systematics introduced with choice will affect future studies investigating the MZR and FMR, especially as we begin establishing these principal scaling relations in the high-\textit{z} Universe \citep[e.g.,][]{Curti_2023_MZR}. The slope and normalization of these scaling relations are essential in constraining galaxy chemical evolution models and interpreting the driving mechanisms behind their respective existence, shape, and evolution, and thus self-consistency is key before the systematics and their effects are more closely examined.

\section{Calibration of the new \^R diagnostic}
\label{appendix:new_calibration}

In Section~\ref{Mirko's New Calibration} we provide the calibration to a new metallicity diagnostics based on a combination of log([OIII]$\lambda 5007$/H$\beta$) and log([OII]$\lambda 3727,29$/H$\beta$) which  differs from the standard R23, and we test it against galaxies with direct metallicities at high-z ($z>2$) from ERO, CEERS, and JADES.
Here, we provide a more detailed description of the calibration sample and rationale.

The sample combines the stacked spectra of SDSS galaxies in bins of log([OIII]$\lambda 5007$/H$\beta$) vs log([OII]$\lambda 3727,29$/H$\beta$) at high metallicity (12+log(O/H)$\gtrsim8.2$) from \cite{Curti_2017} with individual galaxies at intermediate and low metallicities compiled from the literature.
In particular, the latter include $364$ low-metallcity SDSS and blue compact dwarf galaxies from \cite{Izotov_2006}, $41$ galaxies from \cite{Berg_2012}, $18$ galaxies from \cite{Izotov_2019}, $5$ galaxies from \cite{Pustilnik_2020,Pustilnik_2021}, and 
$95$ galaxies from \cite{Nakajima_2022} (and Nakajima, private communication), for a total of 465 low-metallicity objects with $T_e$-based oxygen abundances.

\begin{figure*}
    \centering
    \includegraphics[width=0.40\textwidth]{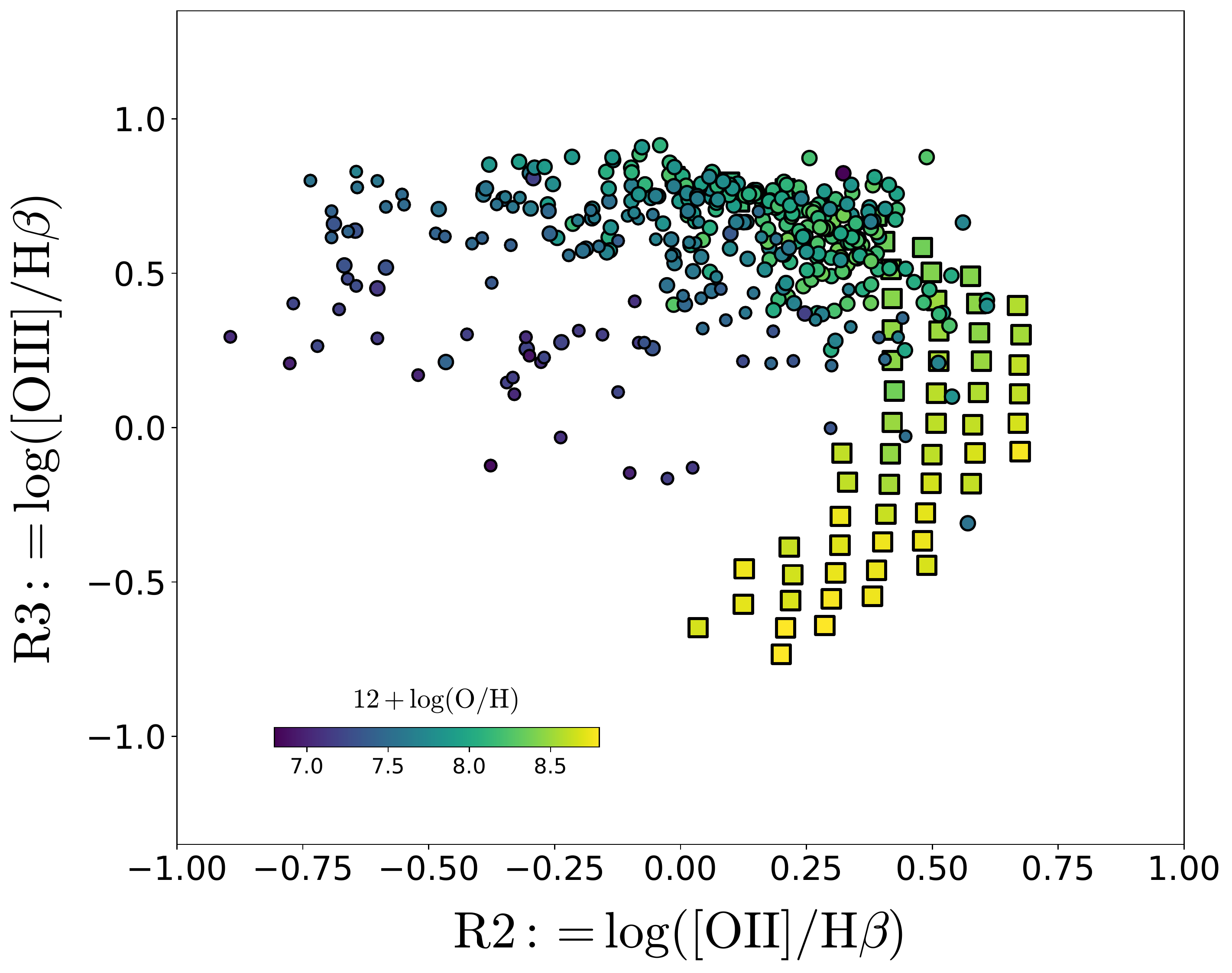}
    \hspace{0.5 cm}\includegraphics[width=0.45\textwidth]{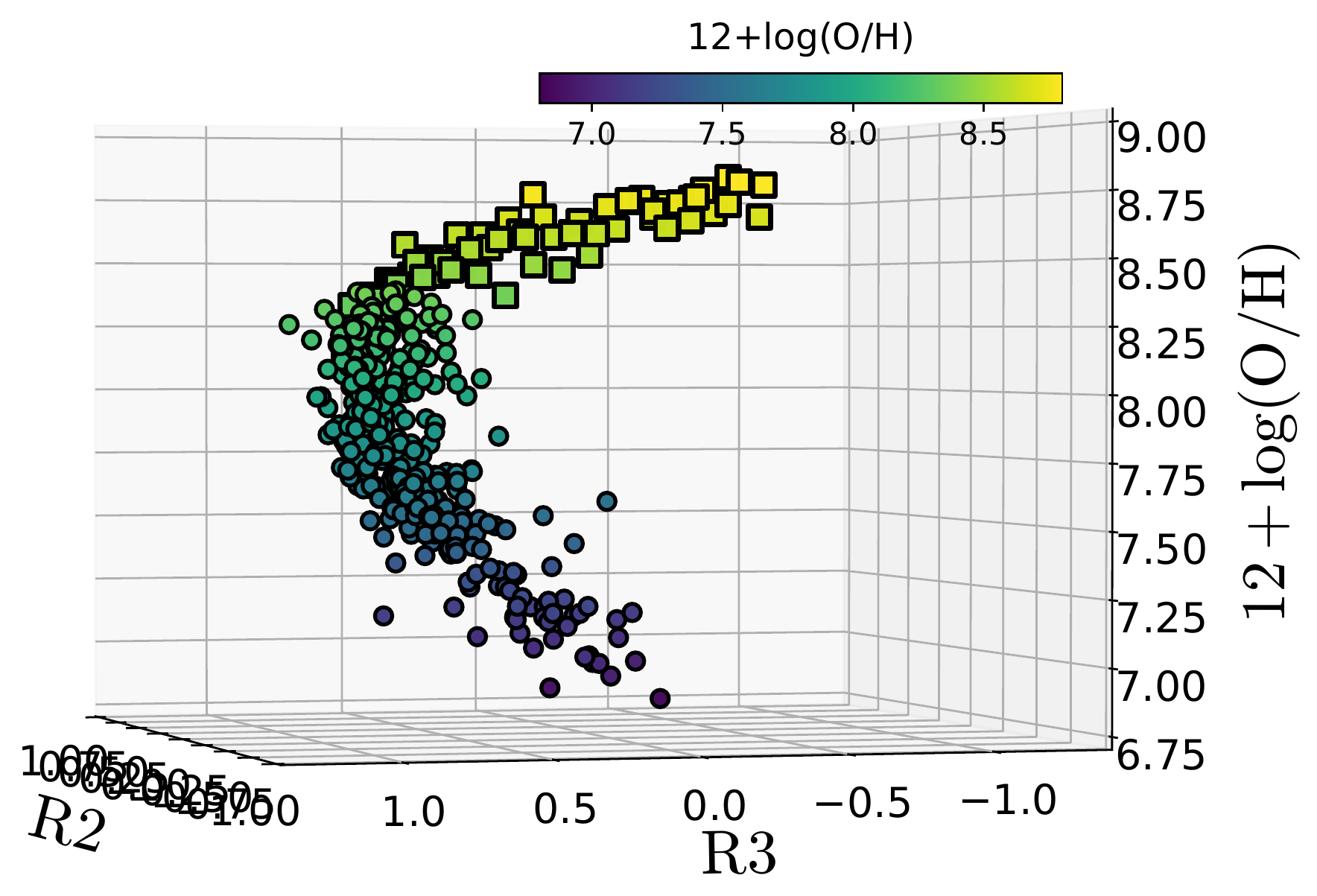} \\
    \vspace{0.75cm}
    \centering  
    \includegraphics[width=0.60\textwidth]{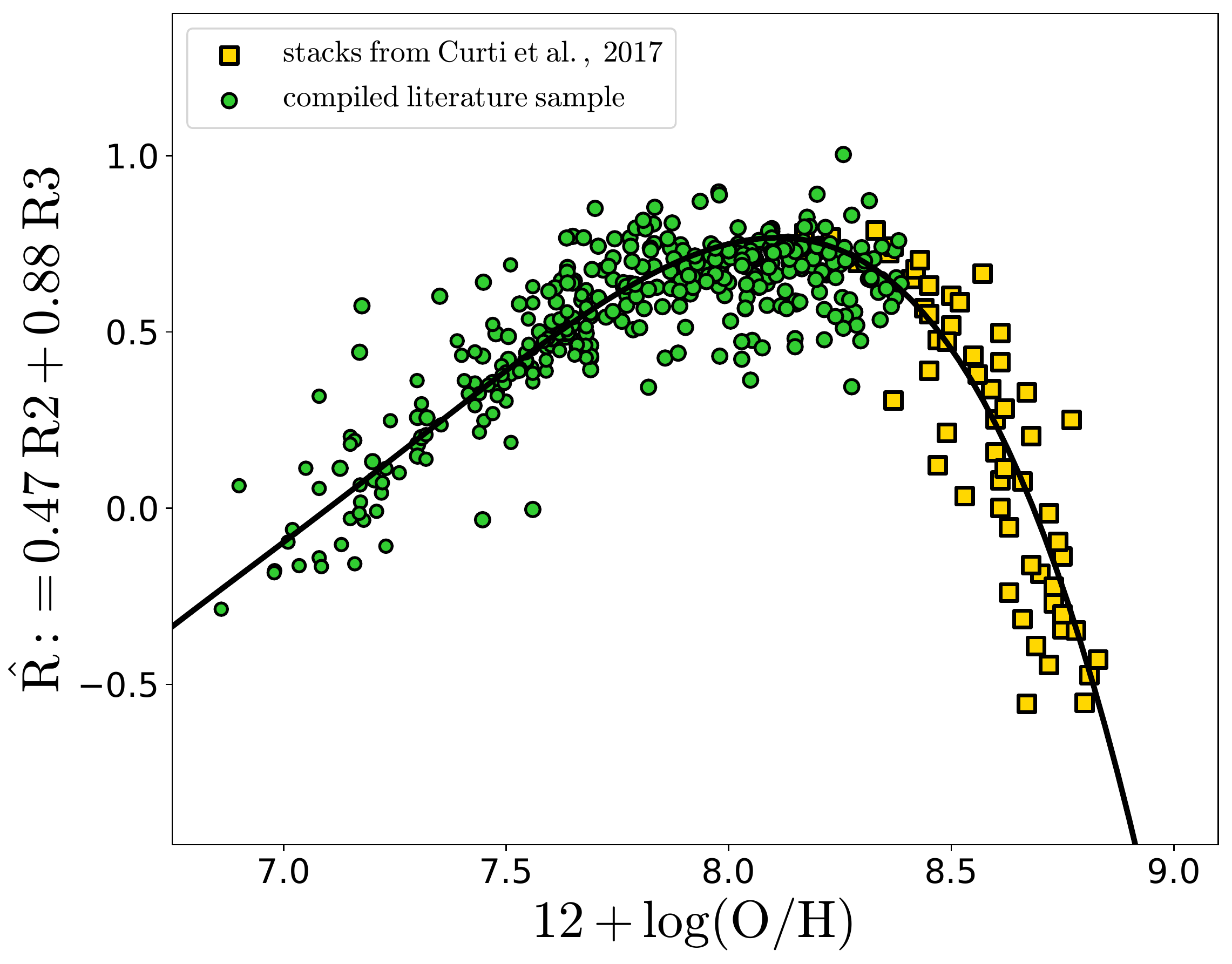}
    \caption{\textit{Top-left panel :} The distribution of our combined sample of stacked spectra (solid squares, from \citealt{Curti_2017}) and individual galaxies (solid circles, compiled from literature as described in the text of Appendix~\ref{appendix:new_calibration}) in the R2 vs R3 diagram. Each point is colour-coded by the Te-derived metallicity. \textit{Top-right panel :} A rotation by $61.82$ degrees of the R2-R3 plane around the O/H axis. Such projection minimise the scatter in metallicity at fixed  \^R = 0.47 R2 + 0.88 R3. \textit{Bottom panel :} The best-fit polynomial relation (black curve) defining the calibration for the \^R diagnostic is shown together with the full calibration sample. }
    \label{fig:new_R_calibration}
\end{figure*}

In the top-left panel of Figure~\ref{fig:new_R_calibration}, we plot the distribution of this sample in the log([OIII]$\lambda 5007$/H$\beta$) vs log([OII]$\lambda 3727,29$/H$\beta$) diagram; each data point is color-coded by its metallicity derived with the $T_e$ method, with squared symbols representing stacked spectra from \cite{Curti_2017} and circles marking individual galaxies from the literature.
The distribution of points in the diagram reflects the well known sequence in metallicity and ionisation parameter observed in large local surveys like SDSS; however, several among the most extremely metal poor galaxies deviate from the sequence in its upper-left branch, while preferentially occupying a region of significantly lower R3, at fixed R2. This makes it difficult to find a parametrisation in such a 2D space that correctly predicts the metallicity over the entire range spanned by the sample. 
 
We therefore search for a re-projection of the axis that facilitate the metallicity prediction over the whole abundance scale. Ideally, such projection should incorporate the different dependence between line ratios, ionisation parameter, and metallicity seen in many metal-poor galaxies of the sample, whose ISM properties more closely resemble those of high redshift objects also observed with JWST/NIRSpec \cite{Cameron_2023}.
The projection is shown in the top-right panel of Figure~\ref{fig:new_R_calibration}.
More specifically, we search for a linear combination of R2 and R3 in the form
\begin{equation}
    \hat{R} = \mathrm{cos(\phi)} R2 +  \mathrm{sin(\phi)} R3 \,
\end{equation}
which is equivalent to a rotation of the R2-R3 plane around the O/H axis. 
We then fit a fourth-order polynomial to the resulting \^R ratio versus the metallicity, in the form of $\text{\^R} = \sum_{n} c_{n}x^{n}$ where $x=\text{12+log(O/H)} - 8.69$ , and indentify the angle $\phi$ that allows to minimize the scatter in metallicity from the best-fit relation.
This procedure leads to a best-fit $\phi = 61.82 \deg$, which translates into \^R $= 0.47 R2 + 0.88 R3$, i.e., the best possible projection of the R2 vs R3 diagram to predict metallicity, given the calibration sample.
The best-fit coefficients for the new \^R calibration are reported below, and the RMS of the fit is $0.058$ dex.
\begin{equation}
    c_{0} = 0.0492 \; ;
    c_{1} = -2.9661 \; ;
    c_{2} = -3.9662 \; ;
    c_{3} = -1.8379 \; ;
    c_{4} = -0.3321
\end{equation}
The new calibration, with its best fit, is shown in the bottom panel of Figure~\ref{fig:new_R_calibration}.

\end{document}